\documentclass[twocolumn, prd, aps, hyphen, nofootinbib, preprintnumbers,
showpacs, superscriptaddress, tightenlines]{revtex4}

\usepackage{amsmath}
\usepackage{amssymb}
\usepackage[english]{babel}
\usepackage{color}
\usepackage{dcolumn}
\usepackage{epsfig}
\usepackage{epstopdf}
\usepackage{graphicx}
\usepackage[latin9]{inputenc}
\usepackage{subfigure}
\usepackage{url}

\usepackage[bookmarks, bookmarksopen, bookmarksnumbered]{hyperref}
\usepackage[all]{hypcap}
\newcommand{\Pv}{{\phi}}
\newcommand{\Wu}{{w}}

\newcommand{\be}{\begin{equation}}
\newcommand{\ee}{\end{equation}}

\newcommand{\ret}{{\mbox{\scriptsize ret}}}

\newcommand{\reg}{{\mbox{\scriptsize R}}}
\newcommand{\sing}{{\mbox{\scriptsize S}}}

\newcommand{\act}{{\text{act}}}
\newcommand{\s}{{\text{S}}}
\newcommand{\vp}{{\varphi}}

\newcommand{\vrr}{{\varphi^\text{R}}}
\newcommand{\tr}{{\tilde\varphi^\text{R}}}
\newcommand{\vs}{{\varphi^\text{S}}}
\newcommand{\ts}{{\tilde\varphi^\text{S}}}

\newcommand{\R}{{\text{R}}}

\newcommand{\eqn}[1]{{Eq.~(\ref{#1})}}
\newcommand{\beq}{\begin{equation}}
\newcommand{\eeq}{\end{equation}}
\newcommand{\calR}{{\mathcal{R}}}

 \newcommand{\bea}{\begin{eqnarray}}
 \newcommand{\eea}{\end{eqnarray}}

\begin{document}

\title{Self-force with (3+1) codes: a primer for numerical relativists}

\author{Ian~Vega}

\affiliation{Institute for Fundamental Theory, Department of Physics,
  University of Florida, Gainesville, FL 32611-8440, USA}
\homepage{http://www.phys.ufl.edu/ift/}

\author{Peter~Diener}

\affiliation{Center for Computation \& Technology,
  Louisiana State University, Baton Rouge, LA 70803, USA}
\homepage{http://www.cct.lsu.edu/}

\affiliation{Department of Physics and Astronomy,
  Louisiana State University, Baton Rouge, LA 70803, USA}
\homepage{http://relativity.phys.lsu.edu/}

\author{Wolfgang~Tichy}

\affiliation{Department of Physics, Florida Atlantic University, Boca Raton,
             FL 33431, USA}

\author{Steven~Detweiler}

\affiliation{Institute for Fundamental Theory, Department of Physics,
  University of Florida, Gainesville, FL 32611-8440, USA}

\begin{abstract}
Prescriptions for numerical self-force calculations have traditionally
been designed for frequency-domain or (1+1) time-domain codes which
employ a mode decomposition to facilitate in carrying out a delicate
regularization scheme. This has prevented self-force analyses
from benefiting from the powerful suite of tools developed and used by
numerical relativists for simulations of the evolution of
comparable-mass black hole binaries. In this work, we revisit a
previously-introduced (3+1) method for self-force calculations, and
demonstrate its viability by applying it to the test case of a scalar
charge moving in a circular orbit around a Schwarzschild black hole.
Two (3+1) codes originally developed for numerical relativity
applications were independently employed, and in each we were able to
compute the two independent components of the self-force and the energy flux correctly to within $< 1\%$.
We also demonstrate consistency between $t$-component of the
self-force and the scalar
energy flux. Our results constitute the first successful
calculation of a self-force in a (3+1) framework, and thus open
opportunities for the numerical relativity community in self-force
analyses and the perturbative modeling of extreme-mass-ratio inspirals.
\end{abstract}


\maketitle

\section{Introduction}

A pressing challenge in gravitational wave source modeling is
the inspiral of a solar mass compact object into a supermassive black hole,
better known as an extreme-mass-ratio inspiral or EMRI. These
inspirals result from scattering processes in the star-rich cores of
galaxies, and tend to be highly-eccentric in the strong field
region of the supermassive black hole \cite{HopmanA2005}. The
intricate gravitational waves they emit are believed to be the most
complicated among LISA sources, and their detection and analysis
promise significant science returns for relativistic astrophysics and
general relativity \cite{Amaro-SeoaneETAL2007}. For this to come to
fruition, precise models of their gravitational waves will be necessary.

Immediately confronting this objective are the dramatically different scales
that characterize EMRIs. First, there is the short length scale of the distortion
on the background spacetime made by the compact object, which will need to
be resolved well enough. Then, there is the large length scale of the supermassive
black hole, which sets the distance to the wavezone, where the gravitational
wave signal is to be extracted. Finally, since a typical
EMRI source for LISA will make about $10^4-10^5$ orbits, long-term evolutions
will be required to produce gravitational wave
templates of use to data analysis. These considerations conspire to
make EMRI modeling a difficult problem for numerical relativity.

At some point, the steady advance of computational technology will allow
the numerical relativity community to tackle the full dynamics of this
binary system. In the meantime it seems prudent to develop approximate
schemes that will reliably produce templates of adequate accuracy. One
such scheme takes advantage of the small mass-ratio (say $\mu/M$) and
treats the problem in a perturbative fashion. At lowest order, the system
is but an infinitesimal test mass moving in a black hole spacetime, for
which we know the motion to be geodesic in that spacetime. For such a
case, the mature formalism of black hole perturbation theory is able to
accurately calculate first-order metric perturbations, from which one
infers gravitational waveforms. The foundations for these sorts of
calculations \cite{ReggeWheeler, Zerilli, Teukolsky73} were laid out
beginning over fifty years ago. However, the accuracy requirements of
LISA, particularly on the phase of the waveform throughout the entire
mission lifetime, demand that our models go beyond this leading order
case. (An instructive scaling argument can be found in Sec.~11.1 of
\cite{DetReview}). One thus has to consistently take into account
next-to-leading order effects on the motion of the particle and the
waveform. From this perspective, the dynamics of the inspiral are viewed as the
motion of a finite (but small) point mass in a background black hole
spacetime. The goal is then to determine these finite-mass effects on both
the motion of the point mass and the corresponding gravitational waveform.

These effects are attributed to the {\sl self-force}; the dissipative
part of which is the more familiar phenomenon of radiation reaction.
For a point mass moving in flat spacetime, this effect is entirely
local. Only emission at a given instant affects the motion of the particle at that same instant.
In curved spacetime, due to scattering with the curvature,
the motion of the particle is affected by fields it gave rise to in its
causal past. This makes the resulting physics much richer than
of the flat spacetime case. The same scenario also applies
to scalar and electric charged particles moving in curved spacetime.
While often interesting in their own right, these also serve as
useful and technically less-demanding toy problems for testing new
methods and techniques.

The effect of the self-force is a small acceleration on the point mass
resulting in a secular deviation away from what would otherwise have
been geodesic motion in the background spacetime. This self-force will need to be calculated and used to
modify the motion of the point mass as often as is practical throughout
the course of the inspiral. Formal expressions for this self-force given
in terms of an integral over the particle's entire past history have been
ironed out in the literature \cite{MinoST1997, QuinnWald1997,
PoissonLR}, but these are hardly convenient for practical
calculations. The challenge is then to come up
with an efficient way to compute self-forces based on these formal
expressions. Several useful prescriptions have been developed to address
this issue \cite{AndersonW2005,CasalsDOW2009}, but most influential of these is the \textit{mode-sum
prescription} \cite{Barack01, BMNOS02, BarackOri02}, which we shall describe below.

An alternative viewpoint of the self-force scenario is that
instead of a backreacting ``force'' accelerating the point mass in the
background spacetime, the motion of the point mass is really geodesic
motion on the distorted background geometry \cite{DetweilerW2003,DetReview}.
In this framework, the task is to determine the appropriate distorted geometry
upon which to impose geodesic motion (or equivalently, the
correct smooth potentials governing the motion of
scalar and electric charges). The procedure thus
involves first computing the metric perturbation $h_{ab}$
induced by the point mass on the background spacetime metric $g_{ab}$
(which would be divergent at the location of the point mass), and then
appropriately regularizing this metric perturbation to give the
correct smooth perturbation $h^{\text{R}}_{ab}$. The motion of the
point mass will then be geodesic motion in the
perturbed spacetime $g_{ab}+h_{ab}^{\text{R}}$.

This perspective has been useful on both theoretical and practical
fronts. Most notably, it has reconciled the notion of a self-force
with our understanding of the  equivalence principle \cite{DetweilerW2003}.
It has also served as the basis of convenient variants of the
original mode-sum prescription
\cite{DetweilerMW2003,Hikida04a,Hikida04b,Hikida04c,DiazRiveraMWD,HaasPoisson2006,Haas2007,VegaD2008,Det08},
and has thus advanced our understanding of this important
calculational technique. The method presented in this manuscript is
another off-shoot of this alternative viewpoint.

Much progress has been made in the calculation of the
self-force on a charge that moves momentarily along some
prescribed geodesic of the background spacetime. In particular,
the mode-sum paradigm has contributed tremendously to our understanding
of the elements of a self-force calculation and continues
to serve as the conceptual backdrop upon which all other
calculational schemes are to be understood. It has now been employed
to evaluate the self-force or self-force effects
in a variety of contexts -- scalar
\cite{Burko00,DetweilerMW2003, DiazRiveraMWD, Hikida04a, Hikida04b, Hikida04c, HaasPoisson2006,Haas2007}, electromagnetic
\cite{HaasPhD}, and gravitational \cite{Det08,BarackSago2007,SagoBarackDet09} --
for point sources  moving along geodesics in a Schwarzschild background.
Results for the Kerr spacetime are rare. The first calculation of
a self force on a scalar charge moving through this background has
been achieved only recently \cite{WarburtonB2009}. Despite this good progress, however, little
has been done to achieve a  \emph{dynamic} calculation of the
self-force that is then used to implement backreaction
on moving point sources.

One of the reasons for this is that computing a self-force is
a complicated process. A typical mode-sum calculation
first requires a decomposition of the problem
(i.e. fields and sources) into modes with, say, spherical or
spheroidal harmonics. This is done in order to avoid having to
handle the divergence in the physical retarded field numerically.
Each mode component of the retarded field turns out to be finite at the
location of the charge, and thus, numerically accessible. It is the
sum of these modes that diverges. Buried in each mode is the piece that actually contributes to the self-force.
The central insight of the mode-sum prescription is a way
to access this relevant piece, based on an asymptotic
analysis of the divergence in the retarded field. In calculating the
self-force then, each computed mode is appropriately regularized (using an
analytic expression determined by the asymptotic analysis)
to leave the piece that contributes to the self-force. The regularized
pieces are then summed to get the full self-force. Convergence of
this sum is typically slow, going as $\sim 1/l^n$, where $l$ is the
maximum mode number, and $n$ is determined by the degree to which
one analytically characterizes the asymptotic behavior of the
divergent physical fields.

In \cite{VegaD2008} (henceforth
referred to as Paper I), we introduced a method for calculating the self-force
designed principally to obviate the apparent necessity of a decomposition into
modes in order to perform
the regularization of the retarded field.
Through this alternative method, we
proposed a (3+1) approach to self-force calculations, which fits
squarely with the expertise and infrastructure found in the numerical
relativity community. Our technique is reviewed in better detail
below. Its core idea, however, is straightforward: rather than
regularizing the retarded field,
one can instead appropriately regularize the source term in the field
equation from the start, and thus have the evolution codes deal with
sufficiently-differentiable fields and sources that require no
further regularization. In other words, we deal with regularization
not as a post-processing step, but as preliminary work that needs to
be performed before any numerical run. This work consists of
appropriately replacing the delta-function representation of a point mass source
by a regularized effective source. Designing this effective source
for the wave equation can be done in such a way that
a numerical evolution yields a differentiable field whose gradient at the location
of the particle automatically gives the full self-force. In addition,
this resulting regular field is such that it becomes the physical retarded
solution in the wavezone, from which fluxes and the all-important
waveforms can be extracted.

Two important features of this approach stand out. First, the
evolution code never has to deal with divergent quantities (though
the fields and sources are of finite differentiability); and
second, both self-force calculation and waveform extraction are
trivial, with accuracies limited chiefly by the accuracy that
can be provided by the evolution code. It is these two features
that call out to the numerical relativity community at large,
for in effect all that is required for a self-force calculation
is a (3+1) code that can accurately evolve wave equations with
sources of limited differentiability. The ease with
which one calculates a self-force in this approach suggests
no significant impediment to implementing backreaction on
the particle.

The idea of finding a good substitute for the delta function source in the context
of self-force calculations is also being pursued by others in similar ways
\cite{BarackG2007,BarackGS2007,CanizaresS2009}. Barack and Golbourn
\cite{BarackG2007} introduced a technique consisting of an $m$-mode
(azimuthal mode) regularization of the delta function source. Their
regularization is also guided by analysis of the
singular behavior of the retarded field at the location of the
particle. First one solves (2+1) wave equations with
regularized sources, then extracts the contribution to the self-force
due to each azimuthal mode, and finally sums these to get the full
self-force. Their approach is similar to ours in that
it provides a way of representing point sources on a grid,
but in keeping with the general strategy of the original
mode-sum procedure (which was an $l$-mode sum), it
is likely to inherit some of the properties our approach
seeks to overcome.

A new approach to evaluating the retarded field by Ca\~{n}izares and Sopuerta
\cite{CanizaresS2009} splits the problem
into inner and outer domains marked by the location of the point charge. They
situate the point source along the boundary shared by both domains,
and then just impose appropriate jump conditions on the fields that cross
the boundary.
With the benefit of having to deal with only smooth fields, this
method takes advantage of the exponential convergence
of a pseudospectral implementation for evaluating the retarded field.
In calculating a
self-force, however, they are still restricted to performing a mode-sum of
what remains after regularizing the output of their evolution code.
A chief advantage of our method is precisely
the fact that it escapes the requirement of a mode decomposition.

In Paper I, we reported an implementation of our method using a time-domain
(1+1) code to compute the self-force and retarded field in the wavezone. By first
breaking into modes, this implementation ran counter to the very
motivations underlying our technique. This was done, however,
mainly to provide a quick proof-of-principle, and to establish a
more direct connection with more familiar approaches to self-force
calculation.

In this work, we report for the first time on the feasibility of
our technique in its intended setting. As a result, we have achieved
the first calculation of a self-force in a (3+1) framework. Two different
codes \cite{Schnetter06a,Tichy:2006qn} were employed
in the implementation of our method, both originally intended for numerical
relativity applications. As in Paper I, we focus on the simplest possible
strong-field scenario involving a scalar point charge interacting with its
own scalar field while in a circular orbit around a Schwarzschild black hole.
In choosing to deal with this simple case, we also intend for this document to
be a self-force primer and an invitation to numerical relativists who are on the lookout for
new challenges. The insights gained from self-force analyses should prove
useful to those wishing to tackle the extreme-mass-ratio regime of black hole
binaries.

\subsection*{Outline and notation}

Section \ref{sec:mainmethod} describes our formalism for replacing a
point-particle delta-function source in a standard field equation with a
particular abstract effective source. The effect is that not only does
the resulting field equal the usual retarded field in the wave zone, but
also the field is finite and differentiable at the particle. This allows
it to be used directly in calculating the self-force acting on the
particle. Some details regarding the effective source are given in
Sect.~\ref{sec:effective}.

We describe a practical test application for our approach to self-force
computations in Sect.~\ref{sec:testproblem}. This test application has
been previously well studied by the self-force community by using more
traditional self-force techniques which are not particularly adaptable to
numerical relativity \cite{DetweilerMW2003,Hikida04a,Hikida04b,Hikida04c}.

Section \ref{sec:3+1} contains the implementation details of
modifications, with an eye on applications to self-force problems,
of two different previously developed numerical relativity projects.

Details of the results following from the applications of these two codes
to our test are given in Sect.~\ref{sec:results}. We
compare the time and the radial components of the self-force, calculated
by our two independent codes, with each other and with very
accurate (but tediously obtained) well-known frequency-domain results
\cite{DetweilerMW2003,Hikida04a,Hikida04b,Hikida04c}. The
time component of the self-force removes energy from the particle, and we
also check its consistency with the energy flux via radiation
down the black hole and out at infinity.

Section \ref{sec:summary} gives a summary of the apparent strengths and
weaknesses of our effective source method for regularizing
self-force problems. In very general terms we describe how currently
available computer codes might be adapted specifically to self-force problems.

We have three appendices. Appendix~\ref{1dmath} gives a 1+1D example of
applying traditional finite differencing operators to a wave equation
where the source is of limited differentiability. This elucidates the
discussion of convergence.
In Appendix~\ref{app:flux} we derive the
relationship between between the time component of the self-force and the
radiative energy-flux into the black hole and out at infinity, in the case
of a circular orbit and a scalar field. And we describe a very elementary,
illustrative flat-space toy problem in Appendix~\ref{toyproblem} which
demonstrates how a problem involving a delta-function point source
can be transformed into one with a smooth source in a
mathematically precise way.

For our tensor notation, we denote regular four-dimensional space
time-time indices with letters taken from the first third of the alphabet
$a,b,\ldots,h$\;, indices which are purely spatial in character are taken
from the middle third, $i,j,\ldots,q$\; and indices from the last third
$r,s,\ldots,z$\ and also $\theta$ and $\phi$ are associated with particular
coordinate components. The operator $\nabla_a$ is the covariant derivative
operator compatible with the metric at hand. Partial derivatives with
respect to $t$ are denoted $\partial_t$, and with respect to a generic spatial
coordinate by $\partial_i$.

For the Schwarzschild metric, we use a coordinate system introduced by
Eddington \cite{Eddington1924} and commonly known as Kerr-Schild coordinates
to describe a Schwarzschild black hole.

Our use of the 3+1 formalism follows York \cite{York79} in all aspects
except his labels for tensor indices.


\section{Field regularization for a scalar charge}
\label{sec:mainmethod}

In this section we review the discussion of our method found in Paper I.
We shall discuss it for the case of a scalar point charge
moving in curved spacetime. A typical self-force computation first involves
solving the minimally-coupled scalar wave equation with a
point charge $q$ source,
\begin{equation}
 \nabla^a\nabla_a\psi^{\text{ret}} = -4\pi q \int_\gamma \frac{\delta^{(4)}(x-z(\tau))}{\sqrt{-g}}\; d\tau,
 \label{eq:wave}
\end{equation}
for the retarded field $\psi^{\text{ret}}$. Here $\nabla_a$ is the
derivative operator associated with the metric $g_{ab}$ of the background
spacetime and $\gamma$ is the world line of the charge defined by
$z^a(\tau)$ and parameterized by the proper time $\tau$. The physical
solution of the resulting wave equation will be a retarded field that is
singular at the location of the point charge. A formal expression
for the self-force given by
\begin{equation}
F_a(\tau)=q(g_a{}^b+u_au^b)\nabla_b\psi^{\text{ret}}(z(\tau))\label{eq:naive}
\end{equation}
would thus be undefined without a proper regularization prescription.
Early analyses \cite{DeWittB1960, MinoST1997, QuinnWald1997}
were based upon a Hadamard expansion of the Green function, and showed
that for a particle moving along a geodesic the self-force could be
described in terms of the particle interacting only with the ``tail'' part
of $\psi^{\text{ret}}$, which is finite at the particle itself. The mode-sum
prescription is effectively a way of regularizing the righthand side
of \eqn{eq:naive} to retrieve the force due to this tail part.
Later \cite{DetweilerW2003} it was realized that a singular part of the field
$\psi^\s$ which exerts no force on the particle itself could be identified
as an actual solution to \eqn{eq:wave} in a neighborhood of the particle.
A formal description of $\psi^\s$ in terms of parts of the retarded Green's
function is possible, but
generally there is no exact functional description for $\psi^\s$ in a neighborhood of the
particle. Fortunately, as is shown in \cite{DetweilerMW2003}, an intuitively
satisfying description for $\psi^\s$ results from a careful expansion about
the location of the particle:
\beq
  \psi^\s = q/\rho + O(\rho^3/\calR^4) \text{ as } \rho\rightarrow0,
\label{eq:psiS}
\eeq
where $\calR$ is a constant length scale of the background geometry and
$\rho$ is a scalar field which simply satisfies $\rho^2=x^2+y^2+z^2$ in a
very special Minkowski-like locally inertial coordinate system centered
on the particle, first described by Thorne, Hartle and Zhang
\cite{ThorneHartle85, Zhang86} and applied to self-force problems in
Refs.~\cite{Detweiler2001, DetweilerMW2003, DetReview}. A detailed discussion
of these coordinates can be found in \cite{DetweilerMW2003, DetReview} and Appendix A of
Paper I. Not surprisingly
the singular part of the field, which exerts no force on the particle
itself, appears as approximately the Coulomb potential to a local observer
moving with the particle.

Our proposal for solving \eqn{eq:wave}, and determining the self-force
acting back on the particle now appears elementary. First we define
\beq
  \tilde\psi^\s \equiv q/\rho
\eeq
as a specific approximation to $\psi^\s$. By construction, we know that
$\tilde\psi^\s$ is singular at the particle and is $C^\infty$ elsewhere.
Also, within a neighborhood of the world line of the particle
\begin{eqnarray}
 \nabla^a\nabla_a\tilde\psi^\s &=& -4\pi q \int_\gamma \frac{\delta^{(4)}(x-z(\tau))}{\sqrt{-g}}\; d\tau + O(\rho/\calR^4),
\nonumber\\ &&  \text{ as }\rho\rightarrow0  .
 \label{eq:waveS}
\end{eqnarray}

It must be pointed out that for Eqs.~(\ref{eq:psiS}) and (\ref{eq:waveS}) to be valid, the
Thorne-Hartle-Zhang (THZ) coordinates must be known correctly to $O(\rho^4/\calR^3)$.
Knowing the THZ coordinates only to $O(\rho^3/\calR^2)$ would
spoil the remainder in Eq.~(\ref{eq:waveS}) which would then have a direction dependent
discontinuity in the limit as $\rho \rightarrow 0$.
The local coordinate frame must be known precisely enough in terms of the
global coordinates for the Coulomb-like potential to be a good
representation of the local singular field.

Next, we introduce a window function $W$ which is a $C^\infty$ scalar
field with
\beq
  W = 1+O(\rho^4/\calR^4) \text{ as } \rho \rightarrow 0, \label{eq:Wzero}
\eeq
and $W \rightarrow 0$ sufficiently far from the particle, in particular in the
wavezone and at the black hole horizon. The requirement that $W$ approaches
1 this way, i.e.\ $O(\rho^4)$, is explained below.

Finally we define a regular remainder field
\beq
 \psi^\R \equiv \psi^{\text{ret}} - W\tilde\psi^\s
\eeq
which is a solution of
\bea
 \nabla^a\nabla_a\psi^\R & = & - \nabla^a\nabla_a (W\tilde\psi^\s) \nonumber \\
 & &     - 4\pi q \int_\gamma \frac{\delta^{(4)}(x-z(\tau))}{\sqrt{-g}}\; d\tau
\label{eq:waveRdelta}
\eea
from \eqn{eq:wave}. Because of our use of $\tilde{\psi}^\s$ as an
approximation of the full singular field, $\psi^\R$ will be contaminated with $O(\rho^3/\calR^4)$-pieces that are only
$C^2$ at the location of the charge but do not affect the self-force.

The effective source of this equation
\beq
  S_{\text{eff}} \equiv - \nabla^a\nabla_a (W\tilde\psi^\s)
             - 4\pi q \int_\gamma \frac{\delta^{(4)}(x-z(\tau))}{\sqrt{-g}}\; d\tau
\label{eq:Seff}
\eeq
is straightforward to evaluate analytically, and the two terms on the
right hand side have delta-function pieces that precisely cancel at the
location of the charge, leaving a source which behaves as
\beq
  S_{\text{eff}} = O(\rho/\calR^4) \text{ as }
  \rho \rightarrow 0 .
\label{eq:SeffBehave}
\eeq
Thus the effective source $S_{\text{eff}}$ is continuous but not
necessarily  differentiable, $C^0$, at the particle while being $C^\infty$
elsewhere\footnote{With $\rho^2 \equiv x^2+y^2+z^2$, a function which is
$O(\rho^n)$ as $\rho\rightarrow0$, is at least $C^{n-1}$ where $\rho=0$.}.

A solution $\psi^\R$ of
\beq
  \nabla^a\nabla_a\psi^\R = S_{\text{eff}}
\label{eq:waveR}
\eeq
is necessarily $C^2$ at the particle. Its derivative
\begin{align}
\nabla_a\psi^\R &= \nabla_a(\psi^{\text{ret}}-W\tilde{\psi}^\s) - \tilde{\psi}^\s\nabla_aW \nonumber \\
               &= \nabla_a(\psi^{\text{ret}}-\psi^\s) + O(\rho^2/\calR^4)
                   \,\,\,\,\,\,\,\,\,\, \rho \rightarrow 0
\end{align}
provides the approximate self-force acting on the particle
when evaluated at the location of the charge.
It should be clear why the behavior of $W$ is chosen as in Eq.~(\ref{eq:Wzero}):
A window function with this behavior would not spoil the $O(\rho^2/\calR^4)$-error
already incurred by using the $q/\rho$ approximation
for the singular field.
Also, in the wavezone $W$ effectively vanishes
and $\psi^\R$ is then identically $\psi^{\text{ret}}$ and provides both the
waveform as well as any desired flux measured at a large distance.

General covariance dictates that the behavior of $S_{\text{eff}}$ in
\eqn{eq:Seff} may be analyzed in any coordinate system. But, only in the
specific coordinates of Refs.~\cite{ThorneHartle85} and \cite{Zhang86}, or the
THZ coordinates, is it so easily shown \cite{DetweilerMW2003} that the simple
expression for $\psi^\s$ in \eqn{eq:psiS} leads to the $O(\rho/\calR^4)$ behavior
in \eqn{eq:SeffBehave} and then to the $C^2$ nature of the solution $\psi^\R$
of \eqn{eq:waveR}.

Self-consistent dynamics of a scalar charge requires that the self-force
act instantaneously. Thus a simultaneous solution of the coupled equations
\bea
  \nabla^a\nabla_a\psi^\R &=& S_{\text{eff}}(x(\tau),u(\tau))
\label{eq:waveRR}
\\
  m\frac{d u^b}{d\tau} &=& q(g^{bc}+u^bu^c)\nabla_c\psi^{\reg}
\label{eq:sclEOM1}
\label{eq:selfconst}
\eea
evolves $\psi^\R$ while self-consistently moving the charge via the self-force of \eqn{eq:sclEOM1}.

Our method effectively regularizes the field itself rather than the
gradient of the field, and this regularization is implicit in the
construction of the effective source, as opposed to most existing
self-force calculations in which the divergent pieces of the individual
modes are explicitly subtracted out. Once $\psi^\R$ is determined, our
method has no need for any further regularization.
The derivatives of $\psi^\R$ determine the self-force, providing
instantaneous access; while $\psi^\R$ is identical to $\psi^{\text{ret}}$
in the wavezone, allowing direct access to fluxes and waveforms.

The tedious aspects of our method reside primarily in the construction
of the effective source. This is mainly
due to the need for the transformation from THZ coordinates to the background coordinates.
This transformation is a function of the location and four-velocity
of the particle at any given instant. For fully consistent dynamics where
the particle location and four-velocity are constantly being
modified, this transformation will itself be changing. Thus,
this coordinate transformation will unavoidably have to be
determined and then applied numerically.

Once the effective source is appropriately constructed,
the only remaining requirement 
is a code capable of evolving the wave equation with
a $C^0$ source.
\section{Effective Source}
\label{sec:effective}

At the heart of our approach is the use of a convenient regular representation of a
point particle source. We refer to this as the \emph{effective source}.
The two main elements which enter this are (1) the
approximate singular field, $\tilde{\psi}^{\sing}=q/\rho$, whose
explicit form in terms of the chosen background coordinates depend on
the position and four-velocity of the particle, and (2) the window
function, $W$, whose main purpose is to localize the support of the
approximate singular field to within the vicinity of the particle.

In tackling the same physical test application 
as in Paper I, no modifications
of $\tilde{\psi}^{\sing}$ were needed for our (3+1) runs, apart from a
trivial replacement of the background coordinates in which to express the
effective source. (Here we use ingoing Kerr-Schild coordinates as opposed
to the Schwarzschild coordinates of Paper I).

However, for the current implementation, we did seek out a more
adaptable window function. In (1+1), it proved sufficient to use a simple window function having a
Gaussian-like profile in $r$:
\begin{equation}
W(r) = \exp{\left[-\frac{(r-r_{\text{o}})^N}{\sigma^N}\right]},
\label{eq:gausswindow}
\end{equation}
where $r_{\text{o}}$ is the radius of the circular orbit in Schwarzschild
coordinates, while $N$ and $\sigma$ are parameters to be chosen
according to the requirements described in \S \ref{sec:mainmethod}.
It is easily verified that all of these required conditions can be met
for a sufficiently large $N$. In principle, these conditions make it
a reasonable choice regardless of the numerical implementation.
In practice, however, this original effective source
had some properties that could potentially burden certain (3+1)
codes. (Some results from our early runs with the original source
were, in fact, what motivated the construction of a new one.)
Specifically, the
choice of a Gaussian-like window leads to significant
large-amplitude, short-scale $(\sim \sigma)$ structure away from the particle. This was
not an issue in the (1+1) case, where
high $r$-resolution ($\Delta r\sim M/25$) and high angular
resolution (with a spherical harmonic decomposition going to
as high as $L=39$) were practical. Of course, the extra structure
away from the particle need not necessarily be a problem for all
(3+1) codes. One can maintain the Gaussian-like window and simply
adjust its width $\sigma$ to lessen the artificial short-scale structure.
Some of the runs presented below were performed with this original window, using $N=8$, $r_{\text{o}}=10M$ and $\sigma=5.5M$.
The width was chosen in order to make the profile as wide as possible while
still effectively vanishing before the horizon is reached. These runs show
sufficiently good results as well.

Nevertheless, there is merit in using a more flexible window function;
for instance, one with more adjustable parameters that can be tuned to the needs of
any (3+1) code. A convenient choice makes use of the smooth
transition functions introduced in \cite{YunesTOB2006}. Like in Paper I, we have
chosen to apply a window function only along the $r$-direction, in keeping with
the spherical symmetry of the background spacetime.
\begin{widetext}
Consider the smooth transition function
\begin{align}
\label{eq:transition}
 f(x|x_0,&w,q,s)  &= \left\{ \begin{array}{lr}
 0, & x \leq x_0\\
 \dfrac{1}{2}  + \dfrac{1}{2}\tanh \left( \dfrac{s}{\pi} \left\{ \tan\left[\dfrac{\pi}{2w}(x-x_0)\right]
 -\dfrac{q^2}{\tan\left[\dfrac{\pi}{2w}(x-x_0)\right]}\right\}\right)
 , & x_0 < x < x_0+w\\
   1,& x \geq x_0+w.
\end{array}\right.
\end{align}
\end{widetext}
This is a function that smoothly transits from zero to one in the
region $x_0 < x < x_0 + w $. It comes with four adjustable parameters \{$x_0,w,q,s$\}: 
\begin{enumerate}
 \item  $x_0$:  defines where the transition begins.
 \item  $w$: gives the width of the transition region.
 \item  $q$: determines the point $x_{1/2} = x_0 + (2w/\pi) \arctan q$ where the transition function
                 $f(x) = 1/2$ .
 \item  $s$: influences the slope $s(1+q^2)/(2w)$ at $x_{1/2}$ after $w$ and $q$ are chosen.
\end{enumerate}
Using this transition function, a window function for a particle at $r=R$ could be
\begin{equation}
W(r) = \left\{ \begin{array}{lr} f(r|(R-\delta_1-w_1),w_1,q_1,s_1) &  r\leq R  \\ 1 -f(r|(R+\delta_2), w_2,q_2,s_2) & r>R \end{array} \right.
\label{WolfWindow}
\end{equation}
and $W(r) = 1$ in the region $R-\delta_1<r<R+\delta_2$.

This satisfies all of the key requirements for a window function (and more):
\begin{enumerate}
 \item[(a)] $W(R) = 1$;
 \item[(b)]  $d^n W/dr^n|_{r=R} = 0$, for all $n$;
 \item[(c)] $W = 0$  if   $r\in [0, R-\delta_1-w_1] \cup [R+\delta_2+w_2,\infty)$
          (thus making it truly of compact support);
 \item[(d)] and $W = 1$  if $r\in[(R-\delta_1),(R+\delta_2)]$.
\end{enumerate}
For the actual runs that used this window function, we settled on the following choices for these parameters:
$\{\delta_1=\delta_2=0M; q_1= 0.6, q_2=1.2; s_1=3.6, s_2=1.9; w_1=7.9M, w_2=20M\}$.
The inner width $w_1$ was chosen so that the window and effective source go to exactly zero just outside
the event horizon. The rest were picked after extensively looking at many parameter
combinations. The primary criteria were simply that the effective
source would be sufficiently small everywhere and that it did not possess
structure at extremely small scales. A systematic search for the optimal
set of parameters vis-a-vis its effect on self-force accuracy was not
conducted in this study, and is left for future work.

\begin{figure}[htbp]
\centering
\includegraphics[clip,width=9cm,angle=0]{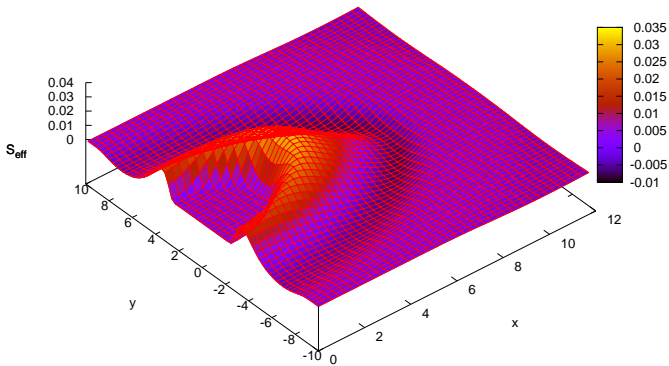}
\caption[Equatorial profile of the new effective source]{Equatorial
profile of the new effective source, $S_{\text{eff}}$, at $\tilde{t}=0$. The axes are
defined simply by $x = r\sin\theta\cos\phi$ and
$y=r\sin\theta\sin\phi$, where $r,\theta,\phi$ are
just the Schwarzschild coordinates (or the Kerr-Schild
coordinates of Sec. V A). The charge in this plot is located
at $X=10M$ and $Y=0$, where the $C^0$ behavior of the source is not
apparent on this scale. Note that much of the structure
induced by the new window function is between the charge and the
event horizon. }\label{newSeff}
\end{figure}

One important attribute of the new window function is that, for a wide range of parameter choices, it leads to an effective source whose over-all
structure away from the particle is significantly less pronounced than
that produced by the original Gaussian-like window.
Comparing the new effective source in Fig.~\ref{newSeff} with the
original source used in Paper I (shown as Fig.~1 of that paper),
one notes immediately that the artificial structure resulting from the new window is almost
two orders of magnitude smaller. Moreover, this structure is mainly located
at $r<R$ (where $R=10M$).

\begin{figure}[htbp]
\centering
\includegraphics[clip,width=9cm,angle=0]{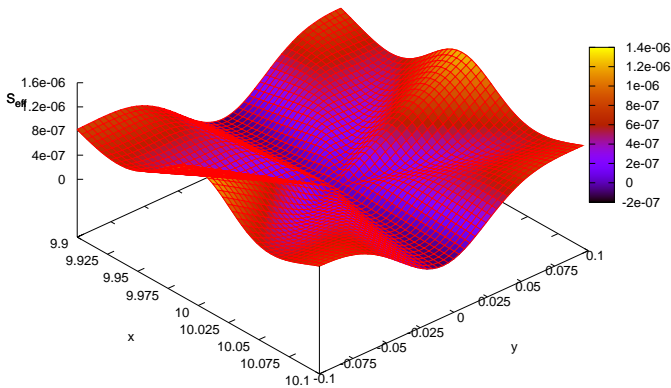}
\caption[$S_{\text{eff}}$ zoomed in at the location of the charge]{$S_{\text{eff}}$ zoomed in at the location of the charge.} \label{newSeffzoom}
\end{figure}

It is instructive to look at the structure of $S_{\text{eff}}$ at the
location of the particle. The effective source, $S_{\text{eff}}$,
is $C^0$ at the particle due to the level of the
approximation used for the singular field $\psi^{\sing}$. This
$C^0$ behavior is sufficient for calculating the
self-force. In our approach, this yields an evolved regular field
$\psi^{\reg}$ that is $C^2$ at the location of the charge, from
which derivatives can be computed to give the self-force.
In Fig.~\ref{newSeffzoom} the $C^0$ nature of the effective
source is revealed.
The effective source is certainly a non-singular
representation of a point charge source which is amenable to the (3+1)
codes we have used for calculating the self-force.

\section{Test Application}
\label{sec:testproblem}
The physical scenario that is analyzed in this paper involves a particle with
mass $m$ and scalar charge $q$ in a perpetual circular orbit around a
Schwarzschild black hole while emitting scalar radiation.
The effects of the self-force (which include the effects of the
emission of radiation) would lead to the gradual decay of the circular
orbit. For simplicity in this analysis we keep the charge in a circular orbit
and compute the external force needed to counteract the scalar
self-force.

With the charge in perpetual circular motion, and
in the absence of other external sources which
may violate this symmetry, the system is helically symmetric.
For any field $G$, there must then exist a helical Killing vector $\xi^a$, such that
\begin{equation}
\pounds_{\xi} G = 0. \label{eq:helixsymm}
\end{equation}
In Schwarzschild coordinates, this Killing vector is simply
\begin{equation}
\xi^a \frac{\partial}{\partial x^a} = \frac{\partial}{\partial t} + \Omega  \frac{\partial}{\partial \phi},
\end{equation}
and
\begin{equation}
\pounds_{\xi}\psi^{\reg} = \xi^a\nabla_a \psi^{\reg} = 0.
\label{killingvect}
\end{equation}
For the circular orbit problem we have chosen, the four-velocity $u^a$ is
tangent to the Killing field at the location of the particle. Thus,
$\nabla_a\phi$ is already orthogonal to $u^a$, and the self-force is given
directly by
\begin{equation}
  F_a = q \nabla_a \psi^\reg
\end{equation}
with no need for the projection operator, present in \eqn{eq:selfconst},
for our test application. 
There are then only two independent components of
the self-force, $F_t$ and $F_r$, with $F_\phi = -F_t/\Omega$ from
\eqn{killingvect}, and $F_{\theta}=0$, by virtue of the system being
reflection symmetric about the equator.

For circular orbits, there exists a useful relation between the scalar
energy flux and $F_t$. In terms of the Kerr-Schild coordinates described
next, this appears as
\begin{equation}
 \left.\frac{dE}{dt}\right|_{r=2M} +
 \left.\frac{dE}{dt}\right|_{r=R}  =
 -\sqrt{1-\frac{3M}{r_{\text{o}}}} F_t,
\label{eq:final_main}
\end{equation}
where
\begin{equation}
\left.\frac{dE}{dt}\right|_{r=2M} = -4M^2\oint
\dot{\psi}^2 \; d\Omega,
\label{eq:fluxH_main}
\end{equation}
and
\begin{align}
 \left.\frac{dE}{dt}\right|_{r=R_\infty} =
 R^2&\sqrt{\frac{R}{R-2M}}\oint_{R} \left[\frac{2M}{R}\dot{\psi}^2 \right.
\nonumber  \\ & + \left.
\left(1-\frac{2M}{R}\right)\dot{\psi}\partial_r\psi\right] \; d\Omega.
\label{eq:fluxR_main}
\end{align}
Here, $r_{\text{o}}$ is the radius of the circular orbit, and $R$ is the finite
outer extraction radius. The field $\psi$ is actually the retarded field, but
one can instead use $\psi^{\reg}$, as long as the surface integrals are
evaluated outside the support of the window function, where (by
design) $\psi^{\text{ret}} = \psi^{\reg}$.
This simple relation is proved explicitly in
Appendix \ref{app:flux}. We use this as a consistency check on our
self-force results.
\subsection{Coordinates for a Schwarzschild black hole}

We describe the Schwarzschild metric as
\begin{equation}
  g_{ab} = \eta_{ab}+ H k_a k_b
\label{eq:KSmetric}
\end{equation}
using a coordinate system first identified by Eddington\footnote{Actually
Eddington \cite{Eddington1924} and Finkelstein \cite{Finkelstein58} wrote
down  the Schwarzschild metric using the precise coordinates of
\eqn{KSschw}. But, somehow the Eddington-Finkelstein duo are associated
with a coordinate system that contains an ingoing or outgoing null
coordinate, although neither explicitly introduced or used such a null
coordinate. While Kerr and Schild (nearly forty years after Eddington)
described the Kerr metric in a form that reduces to \eqn{KSschw} in the
Schwarzschild $a\rightarrow0$ limit. Bowing to current conventions of the
numerical relativity community rather
than to historical accuracy, we label the coordinate system in use as
``Kerr-Schild.''}  and commonly known as Kerr-Schild ingoing coordinates
$(t,x,y,z)$, where $\eta_{ab}$ is the flat Minkowskii metric with
$(-1,1,1,1)$ along the diagonal,
\begin{eqnarray}
 k_a dx^a &=& - dt - \bigg(\frac{x}{r}dx + \frac{y}{r}dy + \frac{z}{r}dz\bigg)
 \nonumber\\
  {}     &=& - dt - dr
\label{KSschw}
\end{eqnarray}
which is the ingoing principle null vector, and
\begin{equation}
 H = \frac{2M}{r}
\end{equation}
with $r^2 = x^2 + y^2 + z^2$.  This is equivalent to the usual Schwarzschild form of the metric
\begin{eqnarray}
  ds^2 &=& -\bigg(1-\frac{2M}{r}\bigg)d\tilde t^2 + \frac{dr^2}{1-2M/r}
  \nonumber \\
  && \hskip.5in {} + r^2(d\theta^2 + \sin^2\theta \,d\phi^{2}) .
\end{eqnarray}
with the Schwarzschild time coordinate $\tilde t$ related to the
Kerr-Schild coordinates by
\begin{eqnarray}
  \tilde t &=& t - 2M\ln(r/2M-1)
\end{eqnarray}
and the usual flat space relationships between $(r,\theta, \phi)$ and $(x,y,z)$.

The Kerr-Schild form of
the metric is popular in the numerical relativity community because a
constant $t$ hypersurface is non-singular and horizon penetrating
which allows for convenient imposition of boundary conditions or for excision.

However, it can be confusing to compare the components of the self-force
as evaluated in these Kerr-Schild coordinates with the components as
evaluated in Schwarzschild coordinates. The Kerr-Schild radial coordinate
$r_{\text{KS}}$ equals the Schwarzschild radial coordinate
$r_{\text{Sch}}$, but constant-$t$ and constant-$\tilde t$ surfaces are
not the same. The relationships between the components of the self-force
for these two coordinates systems are
\begin{align}
  F_{\tilde t}^{\text{Sch}} &= F_t^{\text{KS}}
\label{eq:Ft} \\
  F_r^{\text{Sch}} &= \left(\frac{2M}{r_{\text{o}}-2M}\right)
                F_t^{\text{KS}} + F_r^{\text{KS}}.
\label{eq:Fr}
\end{align}

\subsection{The $3+1$ version of the Schwarzschild metric in
            Kerr-Schild coordinates}

For the Schwarzschild metric in Kerr-Schild coordinates the contravariant form of
the metric (\ref{eq:KSmetric}) is
\begin{equation}
  g^{ab} = \eta^{ab} - H k^a k^b
\label{eq:g^ij}
\end{equation}

With the $3+1$ formalism \cite{York79} the contravariant components of the
metric are closely related to the lapse function $\alpha$, shift
vector $\beta^i$ and spatial metric $\gamma^{ij}$ of a foliation of
spacetime by
\begin{equation}
  g^{ab} =
 \left(
\begin{matrix}
   -\alpha^{-2}   & \beta^j/\alpha^2\\
    \beta^i/\alpha^2  & \quad \gamma^{ij} - \beta^i \beta^j/\alpha^2
\end{matrix}
 \right).
 \label{eq:gAB}
\end{equation}
This relationship gives
\begin{equation}
  -g^{tt}  = 1+H = \alpha^{-2} \,,
\label{eq:gtt}
\end{equation}
\begin{equation}
  g^{it} = H x^i/r = \beta^i/\alpha^{2}  \,,
\end{equation}
and
\begin{equation}
  g^{ij} = \gamma^{ij} - \beta^i \beta^j/\alpha^2
\label{eq:gIJ}
\end{equation}
which implies that
\begin{equation}
  \gamma^{ij} = \eta^{ij} - \frac{H}{1+H} \frac{x^ix^j}{r^2} .
\label{eq:gamIJ}
\end{equation}
Also the determinants of the metrics are related by
\begin{equation}
  \sqrt{-g} = \alpha \sqrt{\gamma}.
\label{eq:rt-g}
\end{equation}

\subsection{The wave equation in Kerr-Schild coordinates}

A form of the wave operator convenient for computation is
\begin{equation}
  \nabla_a \nabla^a \psi = \frac{1}{\sqrt{-g}}\partial_a \Big(\sqrt{-g}g^{ab}
  \partial_b \psi\Big).
\end{equation}
In the $3+1$ formalism, after substitution for the contravariant
components of the metric from Eq.~(\ref{eq:g^ij})
and with the time-independence of $g^{ab}$, we obtain
\begin{eqnarray}
  \alpha^2 \nabla_a \nabla^a \psi
    &=& -\partial_t\partial_t \psi
      + \beta^i \partial_t \partial_i\psi
\nonumber\\ && {}
      + \frac{\alpha}{\sqrt{\gamma}}\partial_i\bigg(\frac{\sqrt{\gamma}}{\alpha}
           \beta^i \partial_t\psi\bigg)
\nonumber\\ && {}
      + \frac{\alpha}{\sqrt{\gamma}} \partial_i\Big[
          \alpha\sqrt{\gamma} \Big( \gamma^{ij}
           -\frac{\beta^i \beta^j}{\alpha^2} \Big) \partial_j\psi\Big]
\label{eq:wave-eq}
\end{eqnarray}
for the wave equation in the Schwarzschild geometry with Kerr-Schild
coordinates and $\alpha$, $\beta^i$ and $\gamma^{ij}$ given in
Eqs.~(\ref{eq:gtt})--(\ref{eq:gamIJ}).

\section{(3+1) Implementations}
\label{sec:3+1}

In the following we describe the finite differencing and pseudospectral
codes used in the numerical experiments.

\subsection{3D multi-block finite difference code}

We solve the wave equation (\ref{eq:waveRR}) for $\psi$
on a fixed Schwarzschild background with a source over a multi-block domain
using high order finite differencing. The code is
described in more detail in~\cite{Schnetter06a}, here we will just summarize
its properties. We use touching blocks, where the finite differencing operators
on each block satisfies a Summation By Parts (SBP) property and where
characteristic information is passed across the block boundaries using penalty
boundary conditions. Both the SBP operators and the penalty boundary conditions
are described in more detail in~\cite{Diener05b1}. The code has been
extensively tested and was used in~\cite{Dorband05a} to perform simulations
of a scalar field interacting with Kerr black holes and was used to extract
very accurate quasinormal mode frequencies.

After the standard 3+1 split, the wave equation is written in first-order in
time, first-order in space form  in terms of the variables $\rho \equiv
\partial_t\psi$ and $\Pv_i \equiv \partial_i \psi$.
The system of equations being integrated is then
\begin{eqnarray}
\partial_t \rho & = &
        \beta^i \partial_i \rho +
           \frac{\alpha}{\sqrt{\gamma}}\partial_i\Big[\alpha\sqrt{\gamma}\Big( g^{ij} \Pv_j
           +  \frac{\beta^i \rho}{\alpha^2} \Big) \Big]
\nonumber \\ && {}
  \qquad     - \alpha^2 S_{\text{eff}},
  \label{eq:sourceterm}
\\
\partial_t \Pv_i & = & \partial_i \rho,
 \label{eq:vdot}
 \\
\partial_t \psi & = & \rho,
  \label{eq:psidot}
\end{eqnarray}
where \eqn{eq:sourceterm} follows from \eqn{eq:wave-eq}
with $g^{ij} \equiv \gamma^{ij}-\alpha^{-2}\beta^i\beta^j$ as in \eqn{eq:gIJ},
and \eqn{eq:vdot} is an elementary consequence of the definition of $\Pv_i$.
The primary dynamical variables are $u = (\rho, \Pv_i)$, while $\psi$ is evolved
via an ordinary differential equation (no spatial derivatives). Across a boundary with unit
normal vector $\xi_i$ the characteristic modes are:
\begin{eqnarray}
w^0_i & = & \Pv_i - \xi^j \Pv_j \xi_i \label{eq:transverse} \\
w^{\pm}  & = & (\beta^i \xi_i\mp\alpha)\rho + g^{ij} \xi_i \Pv_j,
\label{eq:normal}
\end{eqnarray}
Where the speeds of the two transverse modes in \eqn{eq:transverse}
are $\lambda^0 = 0$ and the speeds of the two normal modes in
\eqn{eq:normal} are $\lambda^{\pm}=-\beta^i \xi_i\pm\alpha$.

The only necessary modifications to the code described in~\cite{Schnetter06a},
in order to apply it to the problem at hand, were the addition of the source term in
\eqn{eq:sourceterm} and to add code to interpolate the time derivative $\rho$ and the
spatial derivatives $\Pv_i$ of the scalar field to the location of the particle.

In addition some optimizations were performed. OpenMP pragmas and directives
were added to allow for simultaneous OpenMP and MPI parallelization for better
performance on modern multi-core machines.
Also a load balancing issue arose that could potentially lead to very poor scaling because
$S_{\text{eff}}$ is expensive to calculate only in the spherical shell where it is non-zero.
This issue was solved by adding data structures that were distributed evenly
among all MPI processes, with just the right size and shape to cover the
spherical shell. The source is then evaluated first (all processors working
simultaneously) on this distributed data structure and then copied into the
main 3D grid functions.

\subsubsection{Boundary conditions and initial data}
The simulations below were all performed using the 6-block system,
providing a spherical outer boundary and spherical inner excision boundary
without any coordinate singularities. We use the Schwarzschild solution in
Kerr-Schild coordinates as the background metric for the scalar field
evolution. The inner radius was chosen to be $R_{\text{in}}=1.8M$ and
the outer boundary was chosen to be at $R_{\text{out}}=400M$ in most
cases (it was placed at $R_{\text{out}}=600M$ in a few runs for more
accurate extraction of the fluxes). Since we are using SBP finite
differencing operators we can evaluate the right hand sides for the
evolution equations, i.e.\ $\partial_t u = (\partial_t \rho, \partial_t
\Pv_i)$, even as we approach the outer boundary (using more and more
one-sided stencils). At the outer boundary we then convert both $u$ and
$\partial_t u$ to characteristic variables using
Eqs.~(\ref{eq:transverse}) and (\ref{eq:normal}), i.e.\ we obtain $(w^0_i,
w^{\pm})$ and $(\partial_t w^0_i, \partial_t w^{\pm})$. We only have to
apply a boundary condition to $w^-$, since this is the only incoming mode.
We do this by adding a suitable penalty term (only at the outer boundary)
to $\partial_t w^-$ of the form $T (g - w^-)$, where $T$ is a penalty
parameter that has to be chosen to be consistent with the SBP operator and
the speed of the mode in order to achieve stability (see more details
in~\cite{Schnetter06a}) and $g$ is the desired incoming characteristic
mode. In this case we use $g=0$, i.e.\ zero incoming mode. We then
transform $(\partial_t w^0_i, \partial_t w^+, \partial_t w^-+T (g-w^-))$
back to a new $\partial_t u$ that is used by the time integrator to update
the primary variables. At the inner boundary, the geometry ensures that
all characteristics leave the computational domain; i.e.\ there are no
incoming modes and therefore we do not apply any boundary condition there.

We do not, a priori, know the correct field configuration and start the
simulation with zero scalar field $\psi(t=0)=0$, zero time derivative
$\rho(t=0) = 0$ and zero spatial derivatives $\Pv_i(t=0) = 0$, as if the
scalar charge suddenly materializes at $t=0$. After the system is evolved
for a few orbits, the initial transient has decayed and the system
approaches a helically symmetric end state. We used the 8-4 diagonal norm
SBP operators and added some compatible explicit Kreiss-Oliger dissipation
to all evolved variables.

With this code we have performed runs for a scalar charge on circular orbits
of radius $r_{\text{o}}=10M$ with both the wide Gaussian profile
window ($N=8$ and $\sigma=5.5M$) and the smooth transition function window
($\delta_1=\delta_2=0 M; q_1=0.6, q_2=1.2; s_1=3.6, s_2=1.9; w_1=7.9M,
w_2=20M$).
We find that the extracted self-force is independent of the window
function (as it should be) and that the only difference between the runs is
in the shape and amplitude of the initial scalar wave pulse.

\subsubsection{Convergence}
The convergence of the code has been extensively tested in~\cite{Diener05b1}
where the evolution of a plane wave moving across a spherical grid was used
as a test problem (i.e.\ no source). It was shown that for all implemented
finite differencing order the code was converging at the expected order.
For example for the 8-4 SBP operators used here we found the expected fifth order global
convergence.

As the source is only $C^0$ at the world line of the particle, it is to be
expected that the scalar field will be $C^2$ there while the main evolution
variables $\rho = \partial\psi/\partial t$ and
$\Pv_i = \partial\psi/\partial x^i$ should be $C^1$ on the world line of the
particle and $C^{\infty}$ everywhere else.
With the finite differencing code, there will be some stencils which are
penetrated by the worldline at a particular event. For those stencils, the
finite differencing errors will be affected by the limited
differentiability of both the source and the field at the particle.
We would expect that
any traditional centered finite differencing operator applied to a $C^1$ field
(regardless of order) should then only be first order accurate: the second
derivative is discontinuous across the world line and
so the second order terms in the Taylor expansion of the operator will
not cancel.

Naively one would then expect that the solution for $\rho$ and $\Pv_i$ at the
particle and thus the extracted self-force would only converge to first
order. However, as shown in Appendix~\ref{1dmath} for the wave equation in
1+1D, the errors in $\rho$ in fact converge at second order in the L2-norm for a
$C^0$ source. In the Appendix, it is also shown that the error is of high frequency with the
frequency increasing with resolution. Thus, for our test application
we cannot demonstrate pointwise convergence
for the quantities $\rho$ and $\phi_i$. But we expect that
the amplitude of any
noise generated near the particle location will converge at second order.
We find below that the extracted self-force components at the location of the
particle are indeed noisy, but that the noise converges to zero at second order.

\subsection{The pseudospectral code}

We solve the wave equation (\ref{eq:waveRR}) for $\psi$
on a fixed Schwarzschild background with a source using
pseudospectral techniques.

We use the SGRID code~\cite{Tichy:2006qn,Tichy:2009_1,Tichy:2009_2}
to numerically evolve $\psi$.
This code uses a pseudospectral method in which all
evolved fields are represented by their values at certain
collocation points. From the field values at these points
it is also possible to obtain the coefficients of a spectral
expansion. As in \cite{Tichy:2006qn} and \cite{Tichy:2009_1}
we use standard spherical coordinates with Chebyshev polynomials in the
radial direction and Fourier expansions in both angles.
Within this method it is straightforward to compute
spatial derivatives.
To obtain the results described below
the SGRID code uses at most $3\times 53+2\times161=481$
collocation points in the radial and only $64\times 48$ in the
angular directions. This small number of points makes it so
efficient that it can run on a single PC or laptop.

As in \cite{Tichy:2006qn} we introduce an extra variable
\begin{equation}
\Pi \equiv -\frac{1}{\alpha}\left(\partial_t \psi - \beta^i \partial_i \psi\right)
\label{eq:Pidef}
\end{equation}
in order to obtain a system of equations that is first order in time
\begin{eqnarray}
\partial_t \psi &=& \beta^i \partial_i \psi - \alpha\Pi,
\\
\partial_t \Pi &=& -\frac{1}{\sqrt{\gamma}} \partial_i [\sqrt{\gamma} ( \beta^i \Pi
+ \alpha\gamma^{ij}\partial_j\psi)] + \alpha S_{\text{eff}} ,
\label{psiPi_sysA}
\end{eqnarray}
which results from Eqs.~(\ref{eq:Pidef}) and (\ref{eq:wave-eq}).
For the time integration we use a fourth order accurate Runge-Kutta
scheme. We implement Eq.~(\ref{psiPi_sysA}) in
the code using the equivalent, specific form
\begin{eqnarray}
\partial_t \Pi &=&
     \beta^i\partial_i \Pi - \alpha g^{ij}\partial_i\partial_j\psi
       + \alpha\Gamma^i\partial_i\psi
\nonumber \\
   & & {}  - g^{ij}(\partial_i\psi)\partial_j\alpha
 + \alpha K \Pi  + \alpha S_{\text{eff}} ,
\label{psiPi_sys}
\end{eqnarray}
where $K$ is the trace of the extrinsic curvature of a constant $t$
hypersurface, and $\Gamma^i$ is given in terms of the Christoffel symbols
of the 3-metric as $\Gamma^i = \gamma^{jk} \Gamma^i_{jk}$. In ingoing
Kerr-Schild coordinates, $K$ and $\Gamma^i$ are given by
\begin{eqnarray}
K & = & \frac{1}{(1+H)^{3/2}} \frac{H}{r}\left (1+\frac{3M}{r}\right ), \\
\Gamma^i & = & \frac{1}{(1+H)^2} \frac{H}{r}
          \left (\frac{3}{2}+\frac{4M}{r}\right )\frac{x^i}{r}.
\end{eqnarray}

For the time integration we use a fourth order accurate Runge-Kutta
scheme. As in \cite{Tichy:2006qn} we find that it is possible
to evolve this system in a stable manner
if we use a single spherical domain, which extends from
some inner radius $R_{\text{in}}$ (chosen to be within the black hole horizon)
to a maximum radius $R_{\text{out}}$. In this case one needs no
boundary conditions at $R_{\text{in}}$ since all modes are going into the
hole there and are thus leaving the numerical domain.
At $R_{\text{out}}$ we have both ingoing and outgoing modes. We impose
conditions only on ingoing modes and demand that they vanish.
However, since we need more resolution near the particle
it is advantageous to introduce several adjacent spherical domains.
In that case one also needs boundary conditions to transfer modes
between adjacent domains. We were not able to find
inter-domain boundary conditions
with which we could stably evolve the system~(\ref{psiPi_sys}).
For this reason we introduce the three additional fields
\begin{equation}
\phi_i = \partial_i\psi ,
\end{equation}
and we evolve the system:
\begin{eqnarray}
  \partial_t \psi &=& \beta^i \partial_i \psi - \alpha\Pi
\nonumber \\
  \partial_t \Pi &=&
         \beta^i\partial_i \Pi - \alpha g^{ij}\partial_i\phi_j
         + \alpha\Gamma^i\phi_i
\nonumber \\
     && {}    - g^{ij}\phi_i\partial_j\alpha
         + \alpha K \Pi  + \alpha S_{\text{eff}}
\nonumber \\
  \partial_t \phi_i &=&   \beta^j\partial_j \phi_i + \phi_j\partial_i\beta^j
                      - \alpha\partial_i\Pi - \Pi\partial_i\alpha .
\label{psiPiphi_sys}
\end{eqnarray}
Note that this system is now first order in both space and time
and it can be stably evolved using the methods detailed below.
Also notice that we evolve the Cartesian components of
all fields.
Due to the introduction of the additional fields $\Pi$ and $\phi_i$
our evolution system is now subject to the constraints
\begin{eqnarray}
\partial_t \psi &=& \beta^i \partial_i \psi - \alpha\Pi, \nonumber \\
\phi_i &=& \partial_i\psi .
\label{Pi_phi_constraints}
\end{eqnarray}

\subsubsection{Characteristic modes}

The characteristic modes of the system~(\ref{psiPiphi_sys})
are~\cite{Scheel:2003vs}
\begin{eqnarray}
\Wu^{\pm} &=& \Pi \pm \xi^i\phi_i
\nonumber \\
\Wu^{0}_i &=& \phi_i - \xi^j\phi_j\xi_i
\nonumber \\
\Wu^{\psi} &=& \psi .
\label{modes}
\end{eqnarray}
For our shell boundaries
$\xi_i$ is a spatial outward-pointing unit vector.
The fields $\Wu^{0}_i$ and $\Wu^{\psi}$ have velocity
$-\beta^i$, while $\Wu^{\pm}$ have velocity $-\beta^i \pm \alpha\xi^i$.

\subsubsection{Domain setup, boundary conditions and initial data}

We typically use 4 adjacent spherical shells as our
numerical domains. The innermost shell extends from
$R_{\text{in}}=1.9M$ to $r_{\text{o}}=10M$. The next two inter domain
boundaries are at $18.1M$ and $27.5M$. The outermost
shell extends from $27.5M$ to $R_{\text{out}}=210M$.
The outermost shell always has 161 collocation
points in the radial direction. The inner shells all
have the same number of points. We vary their number between
29 and 53. For simulations that last longer than about
$390M$ we have observed that reflections from the outer boundary
can reach the particle and introduce errors in the
self-force. For this reason we have also performed
simulations where we add an additional outer shell
with 161 radial points that extends from $210M$ to
$R'_{\text{out}}=400M$.
\begin{figure}
\centering
\includegraphics[clip,width=9cm]{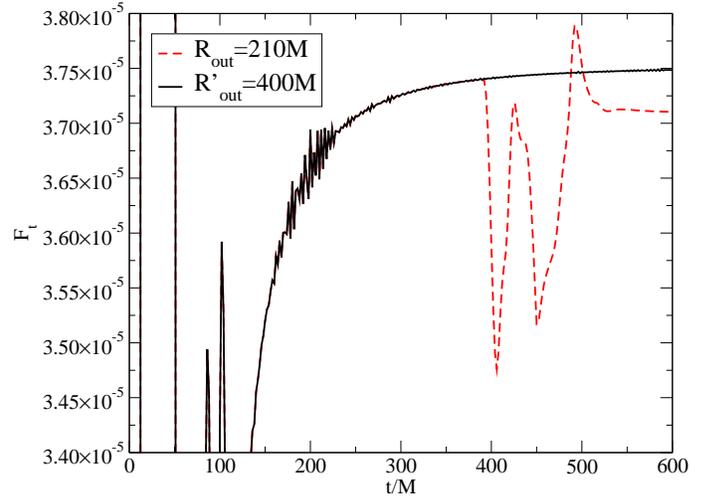}
\caption
{
The broken line shows $F_t$ for an outer boundary located
at $R_{\text{out}}=210M$. We can see reflections arriving
at the location of the particle at around $390M$.
If the outer boundary is moved out to $R'_{\text{out}}=400M$ (solid line)
no such effects can be observed during an evolution time
of $600M$.
}
\label{sgrid_Ft_B210_B400}
\end{figure}
As we can see in Fig. \ref{sgrid_Ft_B210_B400},
we can now evolve to at least $600M$ without spurious
boundary effects.
For our simulations we have used the Window function
in Eq.~(\ref{WolfWindow}).

As mentioned above we do not impose any boundary conditions
at $R_{\text{in}}$. At $R_{\text{out}}$ we impose boundary conditions
in the following way. First we compute $\partial_t \Wu^{+}$ from the fields
$\partial_t\psi$, $\partial_t\Pi$ and $\partial_t\phi_i$ at the boundary.
Then we impose the conditions
\begin{eqnarray}
\partial_t \Wu^{-} &=& -\Pi/r
\nonumber \\
\partial_t \Wu^{0}_i &=& (\delta^k_i - \xi^k \xi_i)\partial_k\partial_t\psi
\nonumber \\
\partial_t \Wu^{\psi} &=& \beta^i\phi_i -\alpha\Pi .
\label{outerBCs}
\end{eqnarray}
on the ingoing modes.
Finally we recompute
$\partial_t\psi$, $\partial_t\Pi$ and $\partial_t\phi_i$
from $\partial_t \Wu^{\pm}$, $\partial_t \Wu^{0}_i$ and $\partial_t \Wu^{\psi}$.
The motivation for the outer boundary conditions in \eqn{outerBCs}
is as follows. The first equation is equivalent to assuming
the Sommerfeld condition $\psi=f(t-r)/r$ for some unknown function $f$.
The other two conditions are derived from the
constraints in \eqn{Pi_phi_constraints}
and can thus be considered constraint preserving.

For the inter domain boundaries we simply compute
$\partial_t \Wu^{\pm}$, $\partial_t \Wu^{0}_i$ and $\partial_t \Wu^{\psi}$
from $\partial_t\psi$, $\partial_t\Pi$ and $\partial_t\phi_i$
at the boundary in each domain. On the left side of the boundary
we then set the values of
the left going modes $\partial_t \Wu^{-}$, $\partial_t \Wu^{0}_i$ and
$\partial_t \Wu^{\psi}$ equal to the values just computed on the right
side of the boundary. On the right side of the boundary we
set $\partial_t \Wu^{+}$ equal to the value computed on the left side.
This algorithm simply transfers all modes in the direction
in which they propagate.

As initial data we simply use $\psi=\Pi=\phi_i=0$.

\subsubsection{Spectral filters}

In order to obtain a stable evolution we apply a filter algorithm
in the angular directions after each evolution step.
As in~\cite{Tichy:2009_1} we project our double Fourier
expansion onto Spherical Harmonics.
After setting the highest $l$ mode in $\psi$ and $\Pi$ to zero
we recompute all fields at the collocation points.
This filter algorithm removes all unphysical modes
and also ensures that $\psi$ and $\Pi$ always
have one less than mode than $\phi_i$.

\subsubsection{Noise reduction}
\label{subsubsec:reduction}

If we compute the coefficients in a Fourier series expansion
of the effective source for a particle moving along a circular orbit
we expect them to be of the form
\begin{equation}
h_m(t) = h_m(0) e^{i m \Omega t} ,
\end{equation}
where $h_m(0)$ are the coefficients at time $t=0$, $m$ is the
mode number and $\Omega$ is the orbital angular velocity.
However, in the SGRID code
we use discrete Fourier transforms instead of Fourier series,
so that the resulting coefficients have a more complicated
time dependence for any finite resolution.

The collocation points in the SGRID code are fixed.
This means that the moving particle periodically approaches
grid points. Thus for any given resolution, the discrete Fourier
coefficients of the effective source will show a modulation
(in addition to the expected phase factor)
on the timescale it takes to move from one grid point to the next.
This modulation is a source of extra noise. In our simulations
we have removed this extra noise by the following procedure.
We simply compute the coefficients $h_m(0)$ once and for all
at $t=0$. For any later time we evaluate the source by
taking the inverse discrete Fourier transform of
$h_m(0) e^{i m \Omega t}$, so that we avoid any extra modulation or
noise.

\subsubsection{Convergence}

\begin{figure}
\centering
\includegraphics[clip,width=9cm]{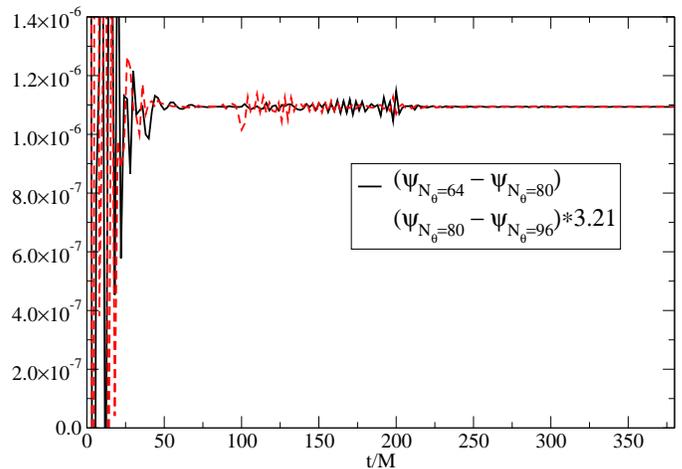}
\caption
{
Differences in $\psi$ at the particle location
for runs with different angular resolutions given by $N_{\theta}$.
All three cases are for $N_r=29$ and $N_{\phi}=3N_{\theta}/4$.
If we scale the difference between medium and high resolutions
by $3.21$, the two curves coincide. This corresponds
to convergence of an order between 4 and 5.
}
\label{sgrid_angconv}
\end{figure}

As the source $S_{\text{eff}}$ is $C^0$ at the particle, we expect
that $\psi$ is $C^2$ and $\phi_i$ is $C^1$ there.
This implies that with our spectral code $\psi$ is expected to be
fourth order convergent at the particle. This expectation is confirmed
by the results presented in Fig.~\ref{sgrid_angconv}.
The solid line shows the difference in $\psi$ between a low
and medium angular resolution run, while the broken line shows
the difference between the medium and high resolution run
scaled by a factor of $s=3.21$ chosen such that the two lines
coincide. This factor is related to the order of convergence $O$ by
\begin{equation}
s = \frac{(1/N_{low})^O - (1/N_{med})^O}{(1/N_{med})^O - (1/N_{hi})^O} .
\end{equation}
For an order of convergence of $O=4$ we would have
\begin{equation}
s = \frac{(1/64)^4 - (1/80)^4}{(1/80)^4 - (1/96)^4} =  2.78 .
\end{equation}
The scale factor of $s=3.21$ thus corresponds to convergence of an
order between 4 and 5.

\section{Self-force and Energy Flux}
\label{sec:results}
In this section we present and then comment on our results. We compute
$F_t$ and $F_r$, the two non-trivial components of the self-force for
a scalar charge in a circular orbit around Schwarzschild, and show
consistency between the results from the two codes.

Using Eqs.~(\ref{eq:fluxH_main}) and (\ref{eq:fluxR_main}), we
also compute the scalar energy flux across the event horizon and
some finite outer boundary, referring to this outer boundary as the extraction
radius. For ease of comparison, these fluxes are expressed as the
$t$-component of the self-force, based on the relation given by Eq.~(\ref{eq:final_main}).

A representative summary of the accuracies we achieved is
presented in Table \ref{finaltable} below.

\begin{table}[h]
\centering
\begin{tabular}{c  c  c  c}
\hline\hline
{} & Code & Result & Error\\
\hline
$F_t$ & mb & $(3.728-3.748)\times 10^{-5}$ & 0.05\%-0.6\% \\
$F_t$ & sgrid  & $(3.7481-3.7487)\times 10^{-5}$ & 0.05\% \\
\hline
$F_r$  & mb & $(1.384-1.389)\times 10^{-5}$ & 0.4\%-0.8\% \\
$F_r$ & sgrid & $(1.384-1.386)\times 10^{-5}$ & 0.4\%-0.5\% \\
\hline\hline
{} & Code & Result & Error\\
\hline
$\dot{E}{(R=150)}$ & mb & $3.773\times 10^{-5}$ & 0.6\% \\
$\dot{E}{(R=150)}$ & sgrid & $3.771\times 10^{-5}$ & 0.6\% \\
$\dot{E}{(R=300)}$ & mb & $3.761\times 10^{-5}$ & 0.2\% \\
$\dot{E}{(R=\infty)}$ & mb & $3.7502\times10^{-5}$  &
0.0005\% \\
\hline\hline
\end{tabular}
\caption[Summary of (3+1) self-force results.]{Summary of (3+1)
results. The top half of the table reports the computed components
of the self-force for a charge in a circular orbit $r_o=10M$. These
were extracted around time, $t$=600M. The error is determined by a
comparison with an accurate frequency-domain calculation
\cite{DetweilerMW2003}, which reports $F_t=3.750227\times10^{-5}$ and
$F_r=1.378448\times 10^{-5}$. The bottom half of the table reports
the computed energy fluxes through the event horizon and the
two-sphere defined by outer extraction radius $R$. The $R=\infty$ case
is an extrapolation to infinite outer extraction radius that was
performed on results of the multi-block code (as explained in
Sec.~\ref{subsection:energyflux}). For ease of comparison with the
local self-force,  all energy fluxes are expressed as $F_t$ according to
Eq.~(\ref{eq:final_main}). ``mb" and ``sgrid" stand for
multi-block and sgrid codes, respectively.}
\label{finaltable}
\end{table}

\subsection{The Dissipative Piece, $F_t$}

The mode-sum of the $t$-component of the self-force,
$\partial_t\psi^{\reg}$, for the case of a scalar charge in a \emph{circular
orbit} in Schwarzschild is known to converge exponentially in $l$, and
is thus typically calculated extremely accurately. Despite the divergence in $\psi^{\text{ret}}$ at the location
of the scalar charge, $\partial_t\psi^{\ret}$ is smooth there and requires no regularization. This arises because the retarded and advanced fields for a charge in
a circular orbit are related by:
\begin{equation}
\partial_t\psi^{\text{ret}} = -\partial_t\psi^{\text{adv}}
\label{eqn:retadv}
\end{equation}

Writing $\psi^{\text{ret}}$ as
\begin{equation}
\psi^{\text{ret}} = \frac{1}{2}(\psi^{\text{ret}}-\psi^{\text{adv}})
+\frac{1}{2}(\psi^{\text{ret}}+\psi^{\text{adv}}),
\end{equation}
we see that the time derivative of the second term vanishes. The first
term is clearly smooth at the location of the charge because it is a
solution of the homogeneous wave equation. For generic orbits,
Eq.~\ref{eqn:retadv} will not be true, and all components of
$\partial_a\psi^{\text{ret}}$ will need to be regularized.

By instantaneously switching on our source at $t=0$, the early part of
the evolution will be contaminated by initial data effects. After some time these transient
effects propagate out of the numerical domain and the system settles down to its helically symmetric end state.

\begin{figure}[htbp]
\centering
\includegraphics[clip,width=9cm,angle=0]{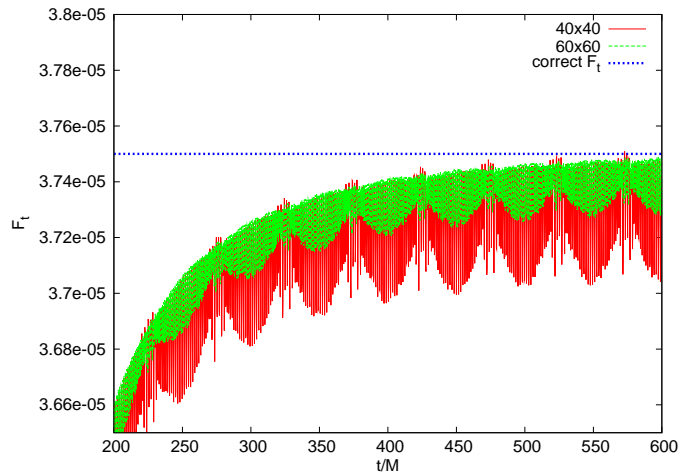}
\caption[$F_t$ computed at different angular resolutions of the
multi-block code]{$F_t$ computed at different angular resolutions of
the multi-block code. The resolutions are described by the number of
angular gridpoints per patch, so that $40\times 40$ corresponds to 40
gridpoints in both $\theta$ and $\phi$ directions within each patch.
The noise in this plot is due to the $C^0$ nature of the effective
source. The frequency of this noise corresponds to the particle
travel time from one gridpoint to the next, and the low-frequency
modulation corresponds to the particle travel time between patch
boundaries.}
\label{FtPeter}
\end{figure}

The results of Fig.~\ref{FtPeter} are from the multi-block code. It
plots $F_t$ evaluated at the location of the charge as a function of
time $t$. Helical symmetry would correspond to a horizontal line,
and we see that the plot gradually approaches this while also getting
to the correct self-force (based on highly-accurate frequency-domain
results in the literature). The particle makes about two full orbits
($T_{\text{orb}} = 2\pi\sqrt{R^3/M} \approx 200M$) before $F_t$ is
reached to within $1\%$. The result improves as initial data
effects further diminish.

On top of this evolution towards the correct self-force appears to be
some sort of modulated noise. This behavior is the result of two
factors. The high-frequency component is
due to the fact that the source is only $C^0$ (and hence derivatives of the
scalar field are only $C^1$) at the location of the particle.
The finite differencing scheme employed here uses stencils near the particle
location that enclose this non-smoothness, and this is expected to introduce
some noise. The frequency of this noise corresponds exactly to the
particle travel time from one grid point to the next. The low-frequency modulation
that envelopes the noise has a period of about $50M$,
which is exactly the time between crossings of inter-patch boundaries.
This is due to the inflated sphere coordinates used within the individual
blocks. The angular resolution is slightly higher near the edges than in
the middle of the block.

We observe that the amplitude of this noise decreases with
increased angular resolution. At $40\times 40$ angular resolution, we obtain
values for the self-force between $3.7\times 10^{-5}$ and $3.75\times 10^{-5}$ i.e.\ within $1.3\%$
of the frequency domain value (which we will consider in the following to be
exact). At an angular resolution of $60\times 60$ the
amplitude of the noise is smaller by a factor of 2.25 corresponding to
second order convergence and an error of about $0.6\%$ of the exact value.

Both results correspond to a radial resolution of $\Delta r = M/10$.
The inner (excision) boundary was placed at $R_{\text{in}}=1.8M$ and the outer
boundary at $R_{\text{out}}=400M$. Modifying the radial resolution (to $\Delta r =
M/15$) does not significantly impact the amplitude of the noise.


\begin{figure}[htbp]
\centering
\includegraphics[clip,width=9cm,angle=0]{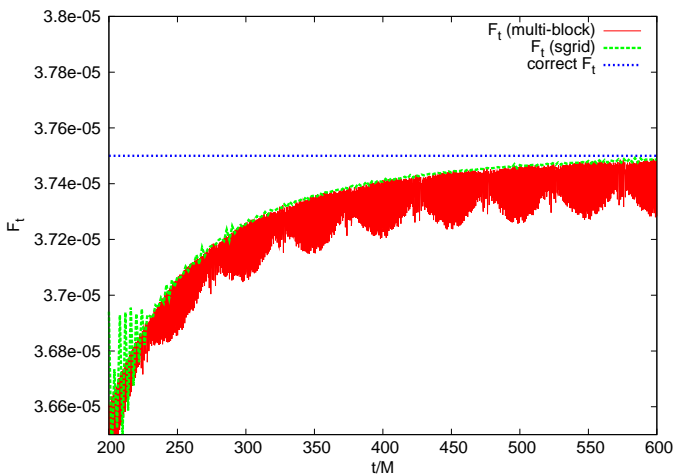}
\caption[Comparing $F_t$ results from the multi-block and SGRID
codes]{Comparing $F_t$ results from the multi-block and SGRID codes.
The multi-block result was achieved with 60$\times$60 angular
resolution and $\Delta r =M/10$ radial resolution, as in Fig.
\ref{FtPeter}. For the SGRID result shown here, the number of
collocation points in the angular directions were $N_{\theta}$ = 64 and
$N_{\phi}=48$. In the $r$-direction, $N_r^{\text{in}}$ =
53, for the three inner shells. The two outer shells were always set to
use $N_r^{\text{out}}$=161 collocation points.}
\label{FtComp}
\end{figure}

In Fig.~\ref{FtComp} we compare the results from the two codes.
There is good agreement between the two, except that the SGRID result
has noticeably less noise than the multi-block result.
This is due to the extra noise reduction performed by the SGRID code,
as described in Sec.~\ref{subsubsec:reduction}.
For the SGRID result shown here, the number of collocation points
in the $\theta$- and $\phi$-directions were $N_{\theta}$ = 64 and
$N_{\phi}=48$, respectively. In the $r$-direction, $N_r^{\text{in}}$ =
53 collocations points were used in each of the three inner shells. (The number of
collocation points in the two outer shells, $N_r^{\text{out}}$, was
kept the same in all runs at $N_r^{\text{out}}$=161.)

\subsection{The Conservative Piece, $F_r$}
The conservative piece of the self-force is really the crucial quantity to compute.
This is the part of the self-force that cannot be inferred from
observations far away
(unlike $F_t$, for example, which can be determined from the energy
flux by using Eq.~\ref{eq:final_main}).
For the case of circular orbits, this conservative piece shows up
entirely as the $r$-component, $F_r$. In a mode-sum self-force
calculation, this would be the quantity whose mode sum converges as $l^{-n}$,
where $n>1$ is typically a small number depending on the number of
regularization parameters one has access to. In our approach,
calculating $F_r$ (where $r$ is the Schwarzschild radial coordinate)
amounts to taking derivatives of the regular field $\psi^{\reg}$,
which corresponds to taking simple algebraic combinations of the
interpolated values of the evolved fields at the location of the charge.

\begin{figure}[htbp]
\centering
\includegraphics[clip,width=9cm,angle=0]{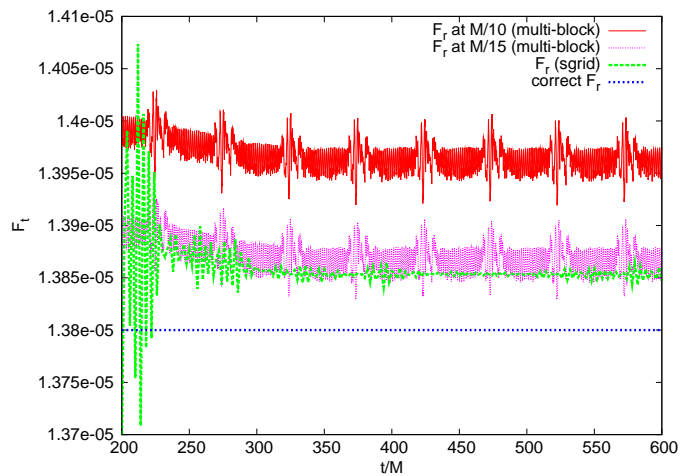}
\caption[Comparing $F_r$ results from the multi-block and SGRID
code]{Comparing $F_r$ results from the multi-block and SGRID codes.}
\label{FrComp}
\end{figure}

We present the results from the two codes together in
Fig.~\ref{FrComp}. The data from the multi-block code were computed using runs at
$40\times40$ angular resolution and two radial resolutions $\Delta r =
M/10, M/15$. For the SGRID
code, we have used data from the same run described in
Fig.~\ref{FtComp}.

We immediately notice that
all the results eventually settle on a value slightly offset from the correct one. It is worth emphasizing though that for the SGRID data and the multi-block data calculated at radial
resolution of $\Delta r = M/15$ the final error is just $< 1\%$.
Moreover, as
can be seen from the two
multi-block results, this offset converges away with increasing radial
resolution.

\subsection{Energy Flux}
\label{subsection:energyflux}

\begin{figure}[htbp]
\centering
\includegraphics[clip,width=9cm,angle=0]{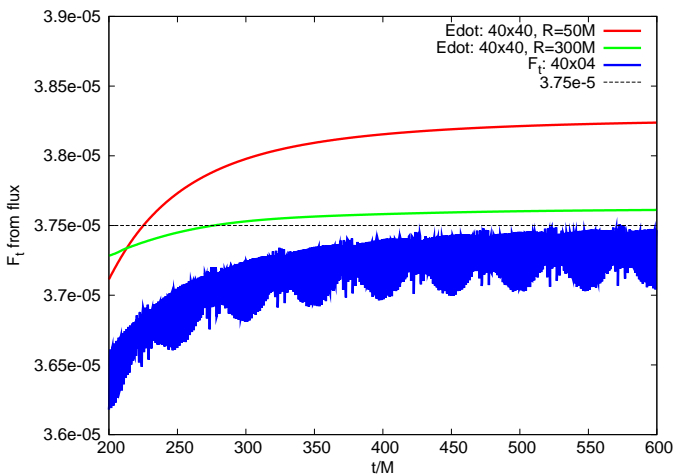}
\caption[Total energy flux computed with the multi-block code]{$\dot{E}$
computed at $40\times 40$ angular resolution of the multi-block code. The
energy flux includes the contribution through the event horizon and
an outer boundary defined by $R$. Shown here are results
from two outer extraction radii, $R=50M$ and
$R=300M$. Energy fluxes are expressed as $F_t$ according
to Eq.~(\ref{eq:final_main}). Also plotted is the local calculation of
$F_t$ at the same resolution.}
\label{FluxPeter}
\end{figure}

\begin{figure}[htbp]
\centering
\includegraphics[clip,width=9cm,angle=0]{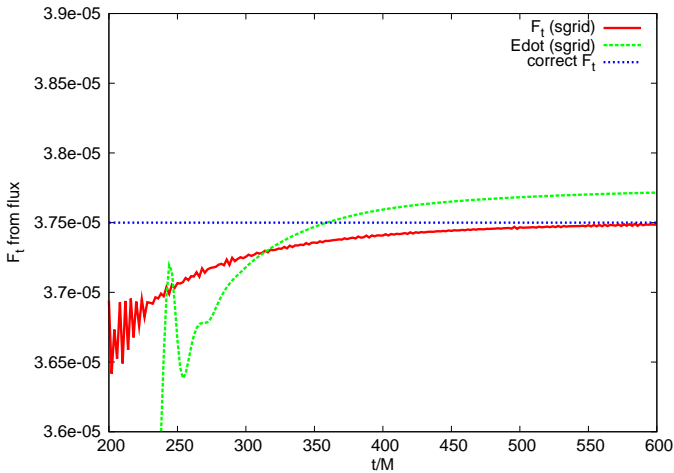}
\caption[Energy flux computed with the SGRID code]{$\dot{E}$ with the
SGRID code from the same run described in Fig.~\ref{FtComp}. This made
use of an outer extraction radius of $R=150M$. Also
plotted is the result of the local $F_t$ calculation.}
\label{FluxAngResWolf}
\end{figure}

An important consistency check for our runs is the relation between
the scalar radiation flux and $F_t$, as given by
Eq.~(\ref{eq:final_main}). Figures \ref{FluxPeter} and
\ref{FluxAngResWolf} display some results from
the multi-block and SGRID codes, respectively.

Figure~\ref{FluxPeter}
features results from the $40\times40$ angular resolution run of
the multi-block code. In this plot, we display the energy fluxes through
two different outer extraction radii, $R=50M$ and $R=300M$, added to
the energy flux through the event horizon. The
outer boundary of the computational domain was at
$R_{\text{out}}=600M$ for both.
For easy comparison, the energy fluxes are converted to a self-force using Eq.~(\ref{eq:final_main}). Also plotted are
the results from the local calculation of $F_t$ (i.e.\ computed by
simply taking the time derivative of the
regular field at the location of the charge) as a function of time
also at the $40\times40$ resolution.
These are all compared with the correct frequency domain result represented by
the straight line.  The flux from the larger extraction radius and the
direct calculation of $F_t$ both show agreement to within $1\%$.
The energy flux is much smoother than the calculated local $F_t$,
since it is an integral over a spherical surface of smooth fields far away from the non-smoothness at the
particle location.

In Fig. \ref{FluxAngResWolf}, we see the corresponding results from
the SGRID code. These come from the same runs described in
Fig.~\ref{FtComp}. The energy flux was calculated
using an outer extraction radius of $R=150M$, and again
converted to the corresponding $F_t$. This is juxtaposed with the
local calculation of $F_t$ and the frequency-domain result. Again, we
observe that except for early-time errors due to spurious initial data,
the energy flux settles to within $1\%$ of the correct answer.

\begin{figure}[htbp]
\centering
\includegraphics[clip,width=9cm,angle=0]{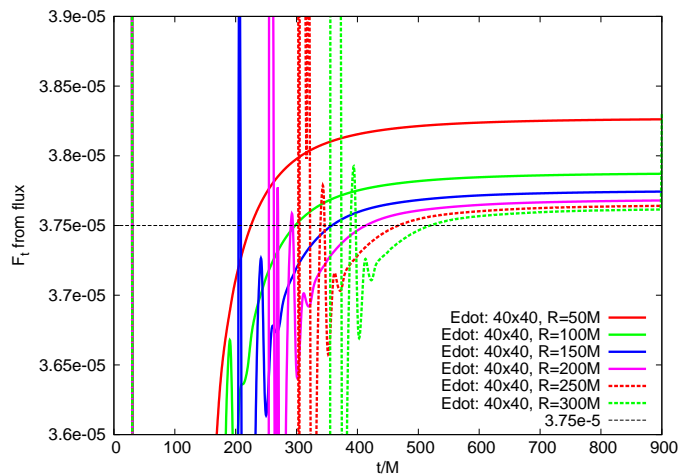}
\caption[Dependence of the (equilibrium) energy flux accuracy on extraction radius]{Dependence of $\dot{E}$
   on various extraction radii. Farther extraction radii is observed to yield better results.}
\label{FluxVersusRadii}
\end{figure}

One notable observation is that the energy flux
improves with increasing extraction radius. This is shown in Fig.~\ref{FluxVersusRadii}. Knowing this, it is
tempting to make the extraction radius as large as possible. However, how far the extraction radius can be
situated is limited by the fact that the flux taken at farther radii naturally takes longer to equilibrate, since the bad initial data waves will have to propagate much farther.


\begin{figure}[htbp]
\centering
\includegraphics[clip,width=9cm,angle=0]{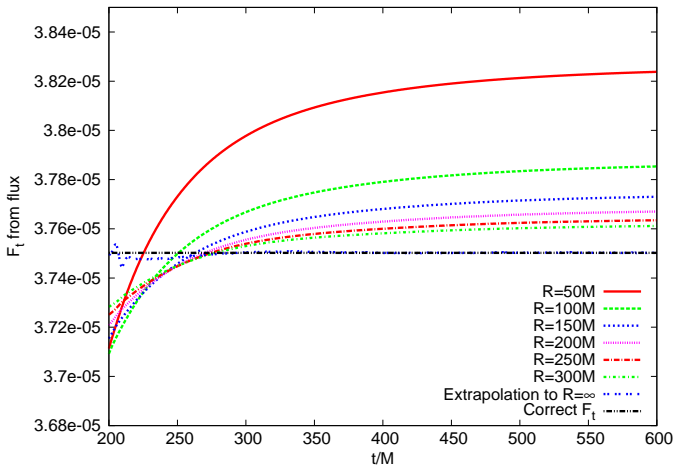}
\caption[Energy flux extrapolation to infinite extraction radius]{$\dot{E}$
    computed with the multi-block code and extrapolated to infinite extraction radius.}
\label{FluxExtrapPeter}
\end{figure}

Instead, one can use results from finite radii to
extrapolate the energy flux in the limit of an infinite
extraction radius. This was done with results from the multi-block
code, and Fig.~\ref{FluxExtrapPeter} shows the outcome. The results
from six finite extraction radii, from $R=50M$ to $300M$, were used to
determine the energy flux in the limit of infinite extraction radius.
First though, one must account for the time shifts in the fluxes. Obviously,
emitted scalar radiation reaches $50M$ first, and only after an interval of time
arrive at the next extraction radius at $R=100M$.

Using the fact that outgoing null geodesics travel at coordinate speed
$(r-2M)/(r+2M)$ in Kerr-Schild coordinates, one can integrate and find
that the time delay between the arrival at various radii are given in
Tab.~\ref{tab:lag}.
\begin{table}[h]
\centering
\begin{tabular}{c  c  c  c}
\hline\hline
Interval & Time Delay \\
\hline
$50M-100M$ & $47.0351M$ \\
$50M-150M$ & $95.7095M$  \\
$50M-200M$ & $144.572M$ \\
$50M-250M$ & $193.687M$ \\
$50M-300M$ & $242.963M$ \\
\hline\hline
\end{tabular}
\caption[Time lags.]{Time lags.}
\label{tab:lag}
\end{table}

Shifting the data by these appropriate time delays, we assume a form for the flux $\dot{E}(R)$ at finite
extraction radius $R$ given by:
\begin{equation}
\dot{E}(R) = \dot{E}(\infty) + \sum_{n=1}^{N} \frac{C_n}{R^n}+O(1/R^{N+1}),
\label{eq:fluxextrap}
\end{equation}
Truncating at $N=5$ the constants $C_1, \ldots, C_N$ and $\dot{E}(\infty)$ can
then be determined from the the six sets of flux data at the
different extraction radii. The resulting $\dot{E}(\infty)$ from this
procedure is
plotted in Fig.~\ref{FluxExtrapPeter}. For reference, we also
include the frequency domain result for $F_t$ expressed as a flux with
\eqn{eq:final_main}. As expected, the agreement is significantly improved.
Extrapolating to infinite
extraction radius, the flux matches to $\sim 0.0005\%$. We take this result
as further validation that our effective source is a good $C^0$
representation for a point charge that would otherwise have been
represented with a delta function.

\section{Summary and Discussion}
\label{sec:summary}
For our flux values to achieve the reported accuracies is noteworthy. This indicates
that our effective source $S_{\text{eff}}$ is a good representation for point particles, in place of
delta functions that are difficult to handle on a grid. Narrow
Gaussian functions centered around the location of the particle have
previously been employed for this task. Reference \cite{Khanna2003} uses a time-domain calculation of the gravitational
energy flux for a point mass orbiting a Kerr black hole which results in errors $\sim 10\%$.
But by optimizing the number of grid points used to sample the narrow Gaussian one can actually
get results of $\sim 1\%$ \cite{Khanna2007}. Recent work by Sundararajan et al.\
\cite{Sundararajan2007} has done even better than this, with a novel discrete representation of the
delta function. Errors of $<1\%$ have consistently been achieved with this technique on
time-domain codes that solve the Teukolsky-Sasaki-Nakamura equation. We note
that our flux results with a (3+1) simulation are already at a comparable
accuracy, albeit only for the scalar energy flux. It is difficult to speculate
on how narrow Gaussians and discrete representations will perform in a (3+1)
context.

The main highlight, however, has to be the accuracy of our self-force results, which have errors
$\lesssim 1\%$. Unlike any other self-force calculations
thus far, these values were calculated by merely taking a derivative of the regular field at the
location of the point charge.
These results are promising as a first attempt at doing a (3+1) self-force calculation.

The judicious placement of collocation points by the SGRID code close to the
location of the charge appears to enable it to represent the effective
source better (and to achieve slightly more accurate results) as opposed to
the uniform grid that the multiblock code uses.
This seems to suggest that
devoting more resources to resolving the region around the charge (like what
would be done with adaptive mesh refinement) is the right strategy.

Both codes show convergence with respect to increases in radial and angular resolution. It is
clear that the finite differentiability of the source reduces the order of convergence of the
codes relative to what it would be if one had a smooth source. For instance, exponential
convergence ought to be observed in the SGRID code, and the
non-smoothness of the source is significant for the multi-block finite
difference code, where only second order convergence is achieved while much higher
order operators are actually used.
Modifications of either code aimed at treating sources with limited differentiability
would be likely to improve the order of convergence. Such a modification might take the
form of an adjustment to a stencil or a spectral function to anticipate the location of the charge.

An important issue has been made apparent by these initial results. Since the self-force is a
very small quantity, the effects of imperfect boundary conditions become a
cause of concern. In both the SGRID and multiblock codes the outer boundary
conditions were implemented in a way that ignored the back scattering off the curvature
of the spacetime outside the computational domain. Since the
self-force contains a tail effect, any such boundary condition will, when the
boundary comes into causal contact with the particle location, affect the
calculation of the self-force. In practice it will seem like the outer
boundary partially reflects the outgoing waves. In order to avoid such
effects we have to place the outer boundaries far enough out, that they
remain out of causal contact with the particle (or the sphere where the
energy flux is measured) for the duration of the run. This of course makes
the runs more computationally expensive both in terms of memory and
cpu usage and limits the number of orbits that can be simulated.
Accurate self-force analyses require careful treatment of the boundary conditions.
One way to do this would be to use the non-local
radiation boundary conditions developed by Lau~\cite{Lau:2004as} and used in
practice for calculating the metric perturbations of an extreme mass ratio
binary with a 1+1D discontinuous Galerkin code~\cite{Field:2009kk}.

In summary, with this preliminary study, we have demonstrated how it is
possible to compute self-forces with existing (3+1) codes---in fact one
of our implementations runs on a laptop!
Moreover, it has been shown that even
in the (3+1) context, the effective source is a good smeared-out alternative to
standard delta-function representations of point sources. The flux resulting
from the effective source matches that due to a point charge with very good
accuracy. There do remain some questions to be explored, like the benefits
of optimizing the codes, the reduction of the convergence order due to the
finite differentiability of the effective source, and the limitations set by the effects
of boundary reflections. As this is merely a first cut analysis
we shall leave these issues for future work.

A goal of this project is to raise interest within the numerical relativity
community in self-force analyses of the EMRI problem. Thus the \texttt{C++}
code which evaluates the effective source for a delta-function scalar charge
has been placed in the public domain via a website
\verb|http://www.phys.ufl.edu/~det/effsource |. Our initial expectation
is to extend this work by allowing for a generic worldline. Our longer term
goal is to have code for an effective source which represents a point mass
orbiting a rotating black hole. At each step as the project progresses we will continue to
put in the public domain all of our code necessary for evaluating the effective source.

\begin{acknowledgments}


I.V.\ acknowledges  the Nutter Dissertation Fellowship from the University
of Florida and the J.\ Michael Harris Fellowship 
from the Institute for Fundamental Theory (URL: http://www.phys.ufl.edu/ift)
for financial support through the course
of this work. S.D.\ and I.V.\ acknowledge the University of Florida
High-Performance Computing Center (URL: http://hpc.ufl.edu)
for providing computational resources.
This work was supported in part by the
National Science Foundation, grants Nos.\ PHY-0652874 and PHY-0855315 with
Florida Atlantic University (W.T.), grant No.\ PHY-0555484 with the University of Florida
(S.D.\ and I.V.),
and through TeraGrid resources provided by LONI and in
part by LONI through the use of Oliver, Poseidon, Louie, Eric and Painter.
We also employed the resources of the Center for Computation \& Technology at
Louisiana State University, which is supported by funding from the Louisiana
legislature's Information Technology Initiative (P.D.).

\end{acknowledgments}

\appendix

\section{Effect of a $C^0$ source on a finite difference code}
\label{1dmath}

In order to better understand the convergence properties of a finite
difference code for the scalar wave equation with a $C^0$ source, we turned
to the 1+1 D case with unit speed in flat space
\beq
-\frac{\partial^2 \psi}{\partial t^2}+\frac{\partial^2 \psi}{\partial x^2}
= S(t,x).\label{eq:1dwave}
\eeq
Similarly to the 3+1 case we can introduce the additional variables
$\rho =\partial_t \psi$ and $\phi = \partial_x \psi$ and rewrite the
wave equation in first-order in time, first-order in space form
\bea
\partial_t \rho & = & \partial_x \phi - S(t,x) \nonumber \\
\partial_t \phi & = & \partial_x \rho \label{eq:wavesystem} \\
\partial_t \psi & = & \rho. \nonumber
\eea
If $\psi(0,x)=0$, $\partial_t\psi (0,x) =0$ and if $S(t,x)=0$ for $t<0$
the solution to \eqn{eq:1dwave} at a given coordinate $(t,x)$ can be shown
to be given in terms of an integral of the source over the domain of
dependence, i.e.\
\beq
\psi(t,x) = -1/2 \int_0^t \int_{x-t+t'}^{x+t-t'} S(t',x')\; dx'\; dt'.
\eeq
The solutions for $\partial_t \psi(x,t)$ and $\partial_x \psi(x,t)$ are
then given by
\begin{align}
\partial_t \psi(t,x) = -1/2 \int_0^t & (S(t',x+t-t')\nonumber \\
 & +S(t',x-t+t'))\; dt' \label{eq:dtpsi}
\end{align}
and
\begin{align}
\partial_x \psi(t,x) = -1/2 \int_0^t & (S(t',x+t-t')\nonumber \\
 & -S(t',x-t+t'))\; dt'.\label{eq:dxpsi}
\end{align}
These integrals can be evaluated numerically in Mathematica to high accuracy
for any given source thus yielding the exact solution to be compared with
an approximate finite difference solution.

A function of the form
\beq
f(t,x) = \exp\left [-\left (\frac{x-at}{c}\right )^2\right ] \tanh(x-at)
\label{eq:source1}
\eeq
is negative for $x<at$ and positive for $x>at$. Forming
\beq
s(t,x) = \| f(t,x) \| - f(t,x) \label{eq:source2}
\eeq
thus results in a source that is positive for $x<at$ and zero for $x>at$.
This source is then $C^0$ at $x=at$ and $C^{\infty}$ everywhere else.

Solving the system of equations in \eqn{eq:wavesystem} with the source in
\eqn{eq:source1} and \eqn{eq:source2} using fourth order centered
finite differencing and fourth order Runge-Kutta time integrator we can then
use the exact solution in \eqn{eq:dtpsi} to calculate the error in the
numerically evaluated $\rho$. For the specific choice of source
parameters $a=\sqrt{2}/2$ and $c=1.3$ we calculated the solution for 3
different spatial resolutions $\Delta x=(0.1, 0.05, 0.025)$ on the spatial
interval $x\in[-6,6]$. The timestep was $\Delta t=\Delta x/4$.

\begin{figure}[htbp]
\centering
\includegraphics[clip,width=9cm,angle=0]{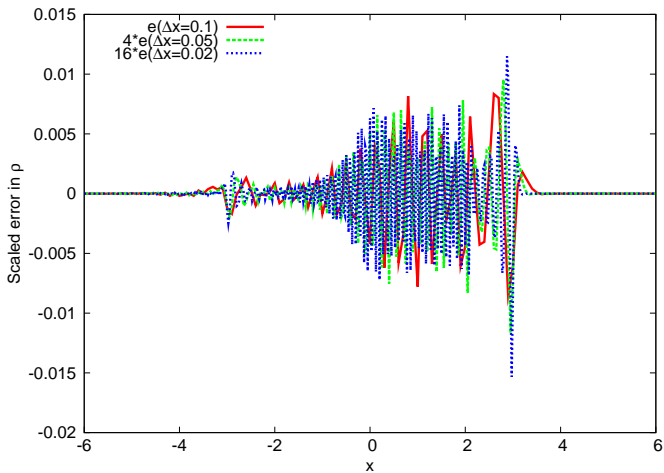}
\caption[Errors in $\rho$]{Scaled errors for second order convergence in
$\rho$ at $t=3$ with a $C^0$ source.} \label{fig:1dwave}
\end{figure}

The scaled errors (for second order convergence) in $\rho$ at $t=3$ can
be seen in Figure~\ref{fig:1dwave}.
As can be seen the errors are as high frequency as can be allowed given the
spatial resolution, i.e.\ the error varies dramatically from grid point to
grid point. Therefore it is impossible to talk about pointwise convergence
since the error at a given gridpoint may be positive at one resolution but
negative at another. However, the amplitude in the error can still be
considered second order convergent. In fact calculating the discrete
L2-norm of the error we find that
$\| e(\Delta x=0.1)\|_2/\| e(\Delta x=0.05)\|_2 = 4.21$ and
$\| e(\Delta x=0.05)\|_2/\| e(\Delta x=0.025)\|_2 = 4.12$ showing that we
have global second order convergence in the L2-norm. The numerical methods
used here are formally fourth order accurate, but because the source is $C^0$
we are limited to only second order convergence.

\begin{figure}[htbp]
\centering
\includegraphics[clip,width=9cm,angle=0]{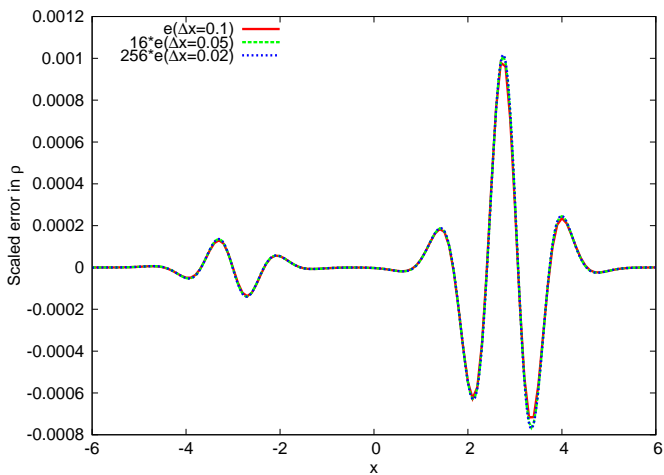}
\caption[Errors in $\rho$]{Scaled errors for fourth order convergence in
$\rho$ at $t=3$ with a $C^{\infty}$ source.} \label{fig:1dwavesmooth}
\end{figure}

If instead we use the source
\beq
S(t,x) = -f(t,x),
\eeq
with the same values for $a$ and $c$ as before we obtain the scaled errors
shown in Figure~\ref{fig:1dwavesmooth}.
Here the errors are smooth and low frequency and the scaled errors from
different resolutions (scaled for fourth order convergence) agree very well, i.e.\
we have pointwise convergence.
For the L2-norms of the errors in this case, we get
$\| e(\Delta x=0.1)\|_2/\| e(\Delta x=0.05)\|_2 = 15.74$ and
$\| e(\Delta x=0.05)\|_2/\| e(\Delta x=0.025)\|_2 = 15.92$, clearly showing the
expected reduction in errors by a factor of 16 with a doubling of the
resolution.

\section{Self-force as boundary integrals in Kerr-Schild coordinates}
\label{app:flux}

In this Appendix, we derive the relationship between the
(Kerr-Schild) time-component of
the self-force on the scalar charge and the energy flux at spatial
infinity and across the event horizon. We also derive explicit expressions for
these fluxes in Kerr-Schild coordinates.

\subsection{How $F_t$ and the energy flux are related for charges in circular orbits}
For a scalar charge going in a circular orbit around a Schwarzschild black hole, there exists a direct
relationship between the time-component of the self-force on the scalar charge and the energy
flux at spatial infinity and across the event horizon.

In the absence of external fields, the motion of a scalar particle is
governed by the self-force acting on it,
\begin{equation}
ma^b = q(g^{bc}+u^bu^c)\nabla_c\psi^{\reg} \equiv F^a.\label{eq:sclEOM}
\end{equation}
The energy per unit mass (i.e.\ specific energy) of a particle along a geodesic with a four-velocity $u^b$
is just $E=-t_bu^b$, where $t^b$ is the time-translation Killing vector of the Schwarzschild spacetime.
The rate of change in this specific energy per unit proper time is $\dot{E}\equiv u^c\nabla_cE
=-u^cu^b\nabla_ct_b - t^bu^c\nabla_cu_b = -t^ba_b$, since $\nabla_{(c}t_{b)} = 0$.
In Kerr-Schild coordinates this is just $\dot{E}=-a_t$.
Evaluating this on a particle moving in a circular orbit,
Eq.~(\ref{eq:sclEOM}) gives us,
\begin{align}
\dot{E}|_p &= -\frac{q}{m}(\partial_t\psi^{\reg} +u_tu^b\nabla_b\psi^{\reg})|_{p}
\\ &= -\frac{q}{m}\partial_t\psi^{\reg}|_{p},
\end{align}
where $p$ signifies the location of the particle.
The second term in the first equality vanishes for a circular orbit,
\begin{align}
u^a\nabla_a\psi^{\reg}|_p &= (u^t\partial_t\psi^{\reg} + u^{\phi}\partial_{\phi}\psi^{\reg})|_p
\\ &= (dt/d\tau)(\partial_t\psi^{\reg} +\Omega\partial_{\phi}\psi^{\reg})|_p \\
&= (dt/d\tau)\pounds_{\xi}\psi^{\reg}|_p \\
&= 0,
\end{align}
where $\xi^a$ is the Killing vector associated with the helical symmetry, so that the third equality
vanishes due to Eq.~(\ref{eq:helixsymm}). Thus
\begin{equation} F_t = -m\dot{E} = q\partial_t\psi^{\reg}. \label{eq:Fteom}\end{equation}
The time component of the self-force in Kerr-Schild coordinates is then just the amount
of energy lost by the particle per unit proper time.

This energy loss must obviously be related to the scalar energy flux. We
shall now derive these relationships and also the explicit expressions for
the energy fluxes in Kerr-Schild coordinates.

The scalar field produced by the charge is determined by the field equation,
\begin{equation}
g^{ab}\nabla_a\nabla_b\psi = -4\pi q\int \frac{\delta^{(4)}(x-z(t))}{\sqrt{-g}}\; d\tau .
\end{equation}
Multiply both sides by $t^a\nabla_a\psi$, integrate over $\mathcal{V}$
(which we take to be the 4-volume bounded by constant Kerr-Schild $t$-surfaces
$t=t_i$ and $t=t_f$, the event horizon, and time-like hypersurface
$r=R$), and $t^a$ is the time-translation Killing vector of
Schwarzschild, and simplify the integral over the delta function to obtain
\begin{align}
\int_{\mathcal{V}} (t^a\nabla_a\psi) & \nabla^b\nabla_b\psi \,\sqrt{-g}\; d^4x
  \nonumber \\ &
          = -4\pi q\int_{t_i}^{t_f}(t^a\nabla_a\psi)|_{p}(dt/d\tau)^{-1}\; dt.
\end{align}
We notice now that the integrand on the left can be expressed as
\begin{eqnarray}
t^a(\nabla_a\psi) \nabla^b\nabla_b\psi & = & t^a g_{ab}\nabla^b\psi \nabla^2\psi \nonumber \\
 & = & 4\pi t^a g_{ab}\nabla_c T^{bc},
\end{eqnarray}
where $T^{bc}$ is the stress-energy tensor for the scalar field,
\begin{equation}
T^{bc} = \frac{1}{4\pi}\left(\nabla^b\psi\nabla^c\psi - \frac{1}{2}g^{bc}\nabla^d\psi\nabla_d\psi\right).
\label{eq:stressenergy}
\end{equation}
We then have
\begin{equation}
\int_{\mathcal{V}} t_a \nabla_c T^{ca} \,\sqrt{-g}\; d^4x
          = - \int_{t_i}^{t_f}F_t(dt/d\tau)^{-1}\; dt,
\end{equation}
where, specializing to circular orbits, $F_t = q\partial_t\psi|_p$.
Here, we have exploited the fact that
$\partial_t\psi^{\text{ret}}$ requires no regularization and thus
equals $\partial_t\psi^{\text{R}}$. Since $t^a$ is a Killing vector, $\nabla_{(c}t_{a)}=0$, and $T^{ca}$ is symmetric in its indices, we have
$t_a\nabla_c T^{ca} = \nabla_c(t_aT^{ca})$. Thus,
\begin{align}
\int_{\mathcal{V}} \nabla_c (t^aT^c{}_{a})\,\sqrt{-g}&\; d^4x \nonumber \\
        & = - \int_{t_i}^{t_f}F_t(dt/d\tau)^{-1}\; dt,
\end{align}
\begin{equation}
\oint_{\partial\mathcal{V}} t^aT_{ca}\; d\Sigma^c
          = - \int_{t_i}^{t_f}F_t(dt/d\tau)^{-1}\; dt,
\label{eq:main}
\end{equation}
where $d\Sigma^c$ is the directional volume element of the
boundary $\partial\mathcal{V}$. The integrand of the left hand side
is essentially the conserved current for the scalar
field, $t^aT_a{}^c$.

We recall again that for the case of a scalar charge in a perpetual
circular orbit of angular velocity $\Omega$,
there exists a helical Killing vector $\xi^a$ given by
\begin{equation}
\xi^a \frac{\partial}{\partial x^a} = \frac{\partial}{\partial t} + \Omega \frac{\partial}{\partial \phi}.
\end{equation}
We break up the left hand side of Eq.~(\ref{eq:main}) into the four
hypersurface integrals,
\begin{widetext}
\begin{equation}
\oint_{\partial\mathcal{V}} t^aT_{ca}\; d\Sigma^c =
\left[\int_{r=2M}-\hat{l}^c r^2\; d\lambda\; d\Omega + \int_{r=R} \hat{r}^c r^2\; dt\; d\Omega
+ \int_{t=t_f} \hat{n}^c \sqrt{h}\; d^3x-\int_{t=t_i} \hat{n}^c \sqrt{h}\; d^3 x \right] t^aT_{ca}.
\end{equation}
\end{widetext}
$\hat{l}^a$ is the null generator of the event horizon, and the rest of the hatted quantities are the respective outward unit normal vectors to
the other hypersurfaces making up $\partial \mathcal{V}$. $\lambda$ is an arbitrary parameter on the null generators $\hat{k}^a$ of the event
horizon. From the helical symmetry of the problem it is easy to see that the last two integrals just cancel each other out. This simply means that
energy content in each constant-$t$ hypersurface is the same. But we can show this explicitly.

Consider the time evolution of the total energy in
a $t$-hypersurface,
\begin{widetext}
\begin{align}
\frac{d}{d t} \int_t\hat{n}^c t^aT_{ac}\sqrt{r(r-2M)}\,\,r^2\; dr&\; d\Omega =
\int_t\frac{\partial}{\partial t}\left(\hat{n}^c t^aT_{ac}\right)\sqrt{r(r-2M)}\,\,r^2\; dr\; d\Omega  \nonumber \\
 &=  - \Omega \int_t \frac{\partial}{\partial \phi} \left(\hat{n}^c t^aT_{ac}\right)\sqrt{r(r-2M)}\,\,r^2\; dr\; d\Omega \nonumber \\
 &=   - \Omega \int\; d \phi \frac{d}{d \phi}
\left(\iint \hat{n}^c
t^aT_{ac}\sqrt{r(r-2M)}\,r^2\sin{\theta}\; dr\; d\theta\right)
= 0,
\end{align}
where we have used the helical symmetry $\pounds_{\xi}F= \left(\frac{\partial}{\partial t}
+ \Omega \frac{\partial}{\partial \phi}\right)F=0$, for any $F$. In other words, time evolution is really just
axial rotation.

Thus, for a circular orbit $r=r_{\text{o}}$, Eq.~(\ref{eq:main}) becomes
\begin{align}
    \left[\int_{r=2M} -\hat{l}^c r^2\; d\lambda\; d\Omega +
    \int_{r=R} \hat{r}^c r^2\; dt\; d\Omega
\right] t^aT_{ca} = -\sqrt{1-\frac{3M}{r_{\text{o}}}}\int_{t_i}^{t_f}F_t\; dt.
\end{align}
\end{widetext}

For convenience, we may set the arbitrary parameter $\lambda$  on the horizon to be $t$. If we then differentiate both
sides with respect to $t$, we finally get
\begin{equation}
 \left.\frac{dE}{dt}\right|_{r=2M} +
 \left.\frac{dE}{dt}\right|_{r=R}  =
 -\sqrt{1-\frac{3M}{r_{\text{o}}}} F_t.
\label{eq:final}
\end{equation}
where
\begin{align}
\left.\frac{dE}{dt}\right|_{r=2M} &= \oint_{r=2M} t^aT_{ca} (-\hat{l}^c) r^2\; d\Omega \label{eq:horizon}, \\
\left.\frac{dE}{dt}\right|_{r=R} &= \oint_{r=R} t^aT_{ca} \hat{r}^c r^2\; d\Omega. \label{eq:infty}
\end{align}

These are the general formulas for the energy flux at spatial infinity and the event horizon. In the next section, we write them
out explicitly in terms of $\psi$ and its derivatives.

Finally, Eqn.~(\ref{eq:final}) can be written in the form,
\begin{equation}
 \left.\frac{dE}{dt}\right|_{r=2M} + \left.\frac{dE}{dt}\right|_{r=R}  = m\frac{dE_p}{dt},
\label{eq:energycons}
\end{equation}
where $E_p=-t_au^a$ is the specific energy of a particle moving along a geodesic. This is just a statement of the conservation of
energy: the energy lost by the charge is also the energy flowing
through $r=2M$ and $r=R$.

\subsection{Scalar energy flux in Kerr-Schild coordinates}
For convenience we write Eqs.~(\ref{eq:horizon}) and (\ref{eq:infty}) in terms of $\psi$ and its derivatives, in Kerr-Schild
coordinates. These formulas are essentially the same except for their unit normals, where one is null and the other spacelike.

We first note that in  Kerr-Schild coordinates, the Schwarzschild metric and its inverse are simply
\begin{align}
g_{ab} &= \eta_{ab} + \frac{2M}{r}k_ak_b, \label{eqn:KSmetric} \\
g^{ab} &= \eta^{ab} - \frac{2M}{r}k^ak^b, \label{eqn:KSmetricinv}
\end{align}
\begin{equation}
k_a = \left(1,\frac{x_i}{r}\right), \,\,\,\,\,\,\, k^a = \left(1,-\frac{x^i}{r}\right)  \label{eqn:nulls},
\end{equation}
where $r^2=x^2+y^2+z^2$ and $\eta_{ab}=\mbox{diag}(-1,1,1,1)$.

We begin first with the energy flux through the event horizon. The event horizon is essentially a surface of constant retarded time
$u = t_{(S)}-r_{(S)}-2M\ln{(r_{(S)}/2M-1)}$, where the subscript $S$ means that these are Schwarzschild coordinates. In Kerr-Schild
coordinates these surfaces of constant $u$ are
\begin{equation}
t=r+4M\ln{(r/2M-1)} + C,
\end{equation}
where $C$ is just a constant. Any particular surface in this family
can be defined parametrically by the equations,
\begin{align}
t &= \lambda, \\
x &= r(\lambda)\sin{\theta}\cos{\phi}, \\
y &= r(\lambda)\sin{\theta}\sin{\phi}, \\
z &= r(\lambda)\cos{\theta},
\end{align}
where $r(\lambda)$ is defined implicitly by the relation
\begin{equation}
\lambda = r+4M\ln{(r/2M-1)}.
\label{eq:rlambda}
\end{equation}
With this, the null generator of the surface (which is also normal to it) is
\begin{equation}
\hat{l}^a \equiv \frac{\partial x^a}{\partial \lambda} = \left(1, \left(\frac{r-2M}{r+2M}\right)\frac{x^i}{r}\right).
\end{equation}

With the stress-energy tensor given by Eq.~(\ref{eq:stressenergy}) and using the expressions given in Eqs.(\ref{eqn:KSmetric})-(\ref{eqn:nulls}),
a small amount of algebra yields
\begin{align}
T_{ab}t^a\hat{l}^b = \dot{\psi}^2 & + \left(\frac{r-2M}{r+2M}\right)\dot{\psi}n^i\partial_i\psi \nonumber \\
 & +\frac{1}{2} \left(\frac{r-2M}{r+2M}\right)\partial_c\psi\partial^c\psi,
\end{align}
where the overdot means a derivative with respect to $t$. At $r=2M$,
the energy flux is then simply just
\begin{equation}
    \left.\frac{dE}{dt}\right|_{r=2M} = -4M^2\oint_{r=2M} \dot{\psi}^2\; d\Omega.
\label{eq:fluxH}
\end{equation}

The normal one-form associated with the hypersurface $r=R$ is
$\xi_a \equiv \partial_ar = (0,x_i/r)$. The corresponding normalized vector is then
\begin{align}
\hat{r}^a &= \sqrt{\frac{r}{r-2M}}\left(\frac{2M}{r},\left(1-\frac{2M}{r}\right)\frac{x^i}{r}\right).
\end{align}
This leads to the following
\begin{align}
    T_{ab}t^a\hat{r}^b = \sqrt{\frac{r}{r-2M}}&\left[\frac{2M}{r} \dot{\psi}^2 \right.\nonumber \\
&\left. + \left(1-\frac{2M}{r}\right)\dot{\psi}\partial_r\psi\right].
\end{align}
Thus, the flux through $r=R$ which is just
\begin{align}
 \left.\frac{dE}{dt}\right|_{r=R} =
 R^2&\sqrt{\frac{R}{R-2M}}\oint_{r=R} \left[\frac{2M}{R}\dot{\psi}^2 \right.
\nonumber  \\ & + \left.
\left(1-\frac{2M}{R}\right)\dot{\psi}\partial_r\psi\right]\; d\Omega.
\label{eq:fluxR}
\end{align}
Taking the limit $r\rightarrow \infty$, this reduces to the more
familiar flat
spacetime case,
\begin{equation}
T_{ab}t^a\hat{r}^b = \dot{\psi}\partial_r\psi.
\end{equation}
And so, we have for the energy flux at spatial infinity,
\begin{align}
\left.\frac{dE}{dt}\right|_{r=\infty} = \lim_{R\rightarrow\infty}R^2 \oint_R  \dot{\psi}\partial_r\psi\; d\Omega.
\end{align}
One of the internal checks we perform is to verify that Eq.~(\ref{eq:final}) holds by computing
the $t$-component of the self-force and the fluxes given in equations (\ref{eq:fluxH}) and (\ref{eq:fluxR}).



\section{A toy illustration of our method}
\label{toyproblem}

Partial differential equations with two dramatically different length
scales are difficult to solve with numerical analysis. Consider the
example of a scalar field $\varphi$ of a spherical object at rest,
centered at $\vec r = 0$ and with a small radius $r_\text{o}$.
And, the small object has a scalar charge density
$\rho(r)$, with $\rho(r)$ being constant for $r \le r_\text{o}$ and
$\rho(r)$ being zero for $r > r_\text{o}$.
 The small object is inside a much larger odd-shaped box, and
$\varphi=0$ on the surface of the box is the Dirichlet boundary condition
for $\varphi$. For simplicity assume that spacetime is flat
and, with the object at rest, there is no radiation and the
field equation is elliptic. Then
\beq
  \nabla^2\varphi = -4\pi \rho \label{del2psi}
\eeq
where $\vec \nabla$ is the usual three-dimensional flat
space gradient operator.

The challenge is to numerically determine the \textit{actual field} $\varphi^\act$
as a function of $\vec r$ everywhere inside the box, subject to the field
equation (\ref{del2psi}) and the boundary condition, and then to find the
total force on the small object which results from its interaction with
$\varphi^\act$.

On the one hand a very fine numerical grid is necessary to resolve $\varphi$ in
and around the object particularly for obtaining the force acting on the
object. On the other hand, a coarse grid would suit the
boundary condition while speeding up the numerical
computation. If the ratio of length scales is many orders of magnitude, or
if the small object is represented by a delta function then adaptive mesh
methods are unlikely to be adequate to resolve the small object
while using modest resources.

To confront the difficulty of the task, we find it advantageous to
introduce the \textit{source field} $\varphi^\text{S}(r)$, where
\begin{eqnarray}
 \text{for} \quad r< r_\text{o}:\quad \varphi^\text{S}(r) &=&
                                 \frac{q}{2r_\text{o}^3}(3r_\text{o}^2-r^2)
\nonumber\\
 \text{for} \quad r> r_\text{o}:\quad \varphi^\text{S}(r) &=& q/r .
\label{defpsiS}
\end{eqnarray}
The source field $\varphi^\text{S}(r)$ is completely determined by local
considerations in the neighborhood of the object, and it is chosen
carefully to be an elementary solution of
\begin{equation}
  \nabla^2 \varphi^\text{S}= -4\pi \rho .
\label{delpsiS}
\end{equation}
Sometimes we call $\varphi^\text{S}$ the \textit{singular field} to
emphasize the $q/r$ behavior outside but near the small source.

The actual scalar field $\varphi^\act$ for the problem at hand is approximately
$\varphi^\s$ near the small object, and the numerical problem may be
reformulated in terms of the \textit{remainder} field
\begin{equation}
  \varphi^\text{R} \equiv \varphi^\act - \varphi^\text{S}
  \label{defpsiR}
\end{equation}
which is then a solution of
\begin{eqnarray}
  \nabla^2 \varphi^\text{R} &=& - \nabla^2\varphi^\text{S}- 4 \pi \rho = 0,
  \label{del2psiR}
\end{eqnarray}
where the second equality follows from \eqn{delpsiS}. The
$\varphi^\text{R}$ is thus a source free solution of the field
equation, and we sometimes call it the \textit{regular field} because the
singular $\varphi^\text{S}$ is removed from the actual field
$\varphi^\act$ in \eqn{defpsiR}.
 And if $\varphi^\R$ is determined then simply adding it to the
analytically known $\varphi^\s$ provides $\varphi^\act$.

A drawback to this formulation might be that the boundary condition requires
that $\varphi^\R = - \varphi^\s$ on the boundary of the box. This is
likely to be more difficult to impose than the original boundary
condition.

A variation of this approach resurrects the original boundary condition. We
introduce a \textit{window function} $W(\vec r)$ that obeys three
properties:
\begin{itemize}
 \item[A.] $W(\vec r) = 1$ in a neighborhood which includes the entire
     source $\rho(r)$, that is all $r<r_{\text{o}}$, but the neighborhood might
     be larger.
 \item[B.] $W(\vec r)=0$ for $r$ greater than some value $r_w$ which
     is not very small.
 \item[C.] $W(\vec r)$ has no structure on the small length scale $r_\text{o}$
\end{itemize}
Then we modify the source field to be $\ts = W(\vec r) \varphi^\s$. Now
the field equation for the regular field $\tilde\varphi^\R$ becomes
\begin{eqnarray}
  \nabla^2 \tilde\varphi^\text{R}
          &=& - \nabla^2\tilde\varphi^\text{S}- 4 \pi \rho = S(\vec r) ,
  \label{del2psiRR}
\end{eqnarray}
and this defines a source $S(\vec r)$ that is zero throughout the small object, is
zero at the boundary and shows no variation on the small length scale $r_\text{o{}}$.
So, as long as $W(\vec r)$ is smooth enough, the
boundary condition for $\tr$ is the natural condition that $\tr = 0$.

In terms of the original source field $\vs$, the source for $\tr$ is
\bea
  S(r) &=& - \nabla^2(W \vp^\s) - 4 \pi \rho
\nonumber\\ &=&
    - \vp^\s\nabla^2W - 2\vec\nabla W\cdot \vec\nabla \vp^\s
             - W\nabla^2\vp^\s  - 4 \pi \rho
\nonumber\\ &=&
    - \vp^\s\nabla^2W - 2\vec\nabla W\cdot \vec\nabla \vp^\s ,
\label{del2phiRPa}
\eea
where the third equality follows from \eqn{delpsiS} and from property A of
the window function.

In the formulation based upon \eqn{del2psiRR}, the small length scale has been
completely removed from the problem.
The field $\tr$ ought to be relatively easy to evaluate, and then the
actual field $\vp^\act= \tr+\ts$ is trivial to determine.

This formulation has the bonus that it simplifies the calculation of the
force on the object from the field. The net force is an integral over the
volume of the object,
\beq
  \vec F = \int \rho(r) \vec \nabla\varphi^\act\; d^3x .
\label{selfF}
\eeq
 In
the original formulation of \eqn{del2psi}, the actual field $\varphi^\act$ in
the integral would be dominated by $\ts$ which changes dramatically over
the length scale of the object, and $\tr$ could be lost easily in the
noise of the computation. The fact that $\ts$ and $\rho$ are spherically
symmetric implies that
\bea
 \int \rho(r) \vec \nabla\ts\; d^3x = 0.
\label{intrhopsiS}
\eea
Then the substitution $\varphi^\act\rightarrow\ts + \tr$ in the integral of
\eqn{selfF} leads to the conclusion that
\beq
  \vec F = \int \rho(r) \vec \nabla\tr\; d^3x .
\eeq
Thus the force acting on the object depends only upon the field $\tr$.

In addition, the field $\varphi^\R$ does not change significantly over a small
length scale, so if the object is extremely small or even a
delta function source then an accurate approximation to the force is
\beq
  \vec F = q \vec \nabla\tr|_{r=0}.
\eeq

This redefinition of the problem at hand is broad enough to encompass a
suggestion by Barack and Golbourn \cite{BarackG2007} to use a window
function that is a step function of unity in an inner region containing the small
object and zero everywhere outside the region. An implementation of
this idea involves solving  for $\varphi^\act$ outside the region, with
the original boundary condition on the box, and solving for
$\vrr$ within the inner region, with the additional boundary conditions that the value and
the normal derivative of $\varphi^\R + \varphi^\s$ match those of $\varphi^\act$
on the boundary of the inner region.

If the source $\rho(r)$ is replaced by a delta function $\delta(\vec r)$
then if $W(\vec r=0)=1$ and $W(\vec r)$ is $C^\infty$ with \textit{all}
derivatives of $W(r)$ vanishing at $\vec r=0$, then $\varphi^\s= q/r$ for all
$\vec r$ and the source $S(\vec r)$ in \eqn{del2phiRPa} is $C^\infty$.
However, if the $n$th derivative of $W$ is \textit{not} zero at $\vec
r=0$, then the source is only $C^{n-4}$. Or, if for some reason the
exact expression for $\vp^\s$ is not known, and only an expansion is available,
then again the source may be of only limited
differentiability.

In applications of this approach to problems in curved spacetime, the
singular field $\varphi^\s$ is rarely known exactly. This limits the
differentiability of the source of \eqn{del2phiRPa} which, in turn,
limits the differentiability of the remainder $\varphi^\R$ at the particle.




\end{document}